\documentclass[bibyear]{aa} 
\usepackage{natbib}

\usepackage{graphicx}
\usepackage[labelfont=bf]{caption}
\usepackage{txfonts}
\usepackage[pdfpagelabels=false]{hyperref}      
\hypersetup{colorlinks=true,linkcolor=blue,citecolor=blue,filecolor=blue,urlcolor=blue,}

\begin{document}

   \title{The [Y/Mg] chemical clock in the Galactic disk \thanks{Based on data obtained from the ESO Science Archive Facility with DOI \url{https://doi.org/10.18727/archive/33}}}

   \subtitle{The influence of metallicity and the Galactic population in the solar neighbourhood}

\titlerunning{[Y/Mg] chemical clock in the Galactic disk}

   \author{J. Shejeelammal, Jorge Mel{\'e}ndez, Anne Rathsam, \and Giulia Martos}

   \authorrunning{J. Shejeelammal et al.}

   \institute{Universidade de São Paulo, Instituto de Astronomia, Geofísica e Ciências Atmosféricas, IAG, Departamento de Astronomia, \\ 
Rua do Matão 1226, Cidade Universitária, 05508-090, SP, Brazil \\
              \email{shejeela@usp.br, jorge.melendez@iag.usp.br}
             }

   \date{Received XXX; accepted YYY}

  \abstract 
   {Stellar ages are an important parameter in studies of the chemical evolution of the Galaxy. To better estimate these ages, various methods complementary to the 
   conventional isochrone fitting method have been implemented in the past decade. Several recent studies have established the existence of a 
   relationship between chemical clocks and stellar ages. The [Y/Mg] clock is a promising technique, but there are still several open questions, 
   such as its validity for metal-poor stars and differences between the thin and thick disk populations.}
   {Our aim is to study the relationship between the [Y/Mg] chemical clock and stellar ages for a sample of solar-type disk stars 
   and to provide the empirical dating relation(s) for the stellar age determination from their precise chemical abundances. 
   {We also studied the effect of metallicity and populations on this chemical clock.} }
   {We derived precise stellar atmospheric parameters as well as the elemental abundances of Mg and Y through line-by-line differential spectroscopic
   analysis for a sample of 48 metal-poor solar-type stars based on high-quality, high-resolution ESO/HARPS spectra. From high-precision \textit{Gaia} astrometric data, 
   stellar masses and ages were estimated through isochrone fitting using Yonsei-Yale isochrones. A joint analysis of our sample, together with
  a sample of 185 solar twins and analogues from our previous works, was performed to calibrate the [Y/Mg] chemical clock in the Galactic disk for $-$0.71 $\leq$ [Fe/H] $<$ +0.34. 
   {Open clusters and stars with asteroseismic ages were used to validate our relations.} }
   {Two different populations are clearly seen in the [Mg/Fe]--[Fe/H] plane: the thick and thin disks. Thick disk stars show an age--metallicity relation, whereas the thin disk 
   shows a flatter age--metallicity distribution. We find a strong, metallicity-dependent anti-correlation between the [Y/Mg] ratio and the stellar ages of our sample. 
   For the first time in the literature, we report similar correlations for thin and thick disk stars.}
   {We find that the [Y/Mg] relation(s) found here for solar-type stars in a wide metallicity range are compatible with those found for solar twins in the literature. 
   Our relation provides high {accuracy and precision (0.45 and 0.99 Gyr, respectively}) comparable with the best accuracy achieved for solar twins to date.}


   \maketitle

\section{Introduction}
In addition to precise chemical abundances, the stellar age is an important parameter in\ studies of Galactic archaeology. 
Combining accurate age estimates with the existing high-precision abundance data can pave the way to a better understanding of
the Galactic chemical evolution \citep{Hawthorn_2016, Alfred_2023}. 
To overcome the problems in the conventional isochrone fitting technique, several complementary 
methods for precisely determining stellar ages have been investigated in the past decade (e.g. \citealt{Soderblom_2010, Howes_2019}).
One such method is the use of chemical clocks -- abundance ratios that show a linear relationship with stellar ages (e.g. \citealt{Nissen_2015, Tuccimaia_2016, Feltzing_2017, Jofre_2020}). 
Also, chemical abundances are being used as part of the training of some machine-learning methods (e.g. \citealt{Li_2022, Moya_2022, Wang_2024}).

\cite{dasilva_2012} provided the first evidence of a linear relation between abundance ratios and stellar ages. One of the most studied
chemical clocks is the [Y/Mg] ratio, and it has been identified as a potential indicator of age in the case of 
solar twins \citep{dasilva_2012, Nissen_2015, Nissen_2016, Tuccimaia_2016, Spina_2016b, Spina_2018}. 
This relation was later found to be valid for solar analogues \citep{Nissen_2017, Nissen_2020}, 
solar-metallicity giants \citep{Slumstrup_2017}, and red clump stars \citep{Casamiquela_2021} 
in the solar neighbourhood. \cite{Skuladottir_2019} explored  the chemical clock outside the Galaxy and found that 
[Y/Mg] can be used as a chemical clock in dwarf galaxies. 

\cite{Feltzing_2017} identified, for the first time, the metallicity dependence of 
the [Y/Mg] clock and noted that it is valid only for [Fe/H] $>$ $-$0.5. The metallicity dependence of the [Y/Mg] chemical clock was later confirmed 
by \cite{Delgadomena_2018, Delgadomena_2019} and \cite{Casali_2020}. \cite{Casali_2020} first determined that the [Y/Mg]--age relation is not universal
and depends on the Galactocentric distance (R$_{GC}$), which was further investigated by \cite{Viscasillas_2022}. This relation has a large dispersion
outside the solar neighbourhood \citep{Casali_2020, Casamiquela_2021}.

The dependence of the chemical clocks on other parameters, such as stellar mass and temperature, was explored by \cite{Delgadomena_2019}. 
The applicability of the [Y/Mg] clock method to thin and thick disk populations of the Galaxy has also been investigated, and different correlations for these two components have been identified \citep{Delgadomena_2018, Titarenko_2019, Tautvaivsiene_2021}. 
These studies have shown that the chemical clock is applicable only to thin disk stars.

Chemical dating can be closely related to the chemical evolution of the Galaxy and its different components (e.g. \citealt{Haywood_2013, Haywood_2016}). 
The aim of this paper is to study the [Y/Mg]--age relation in the solar neighbourhood of the Galactic disk over a wide metallicity range. 
Our analysis is based on a combined sample of 233 solar-type stars with metallicities in the range
$-$0.7 $\leq$ [Fe/H] $<$ +0.34. We examined the metallicity dependence of the [Y/Mg] chemical clock using this sample.  
Few studies in the literature have explored the relation between [Y/Mg] and stellar ages 
for thick disks, and the findings are not always consistent; we investigated this as well. 
This paper is structured as follows: Sect. \ref{section_sample} discusses the stellar sample and the source of high-resolution spectroscopic data. 
In Sect. \ref{section_methodology} we present the spectroscopic analysis. Results are discussed in Sect. \ref{section_results_discussion}.
Finally, concluding remarks are presented in Sect. \ref{section_conclusion}.

\section{Stellar sample and spectroscopic data} \label{section_sample}
{Our analysis is based on a sample that contains three classes of stars of different similarity to the Sun: solar twins, solar analogues, and solar-type (solar-like) stars. 
Solar twins are the stars with parameters $\Delta$T$\rm_{eff}$ = 100 K, $\Delta$log g = 0.1 dex, and $\Delta$[Fe/H] = 0.1 dex of the solar values \citep{Melendez_2006, Ramirez_2009}. 
Solar-analogues are the dwarf stars with parameters $\Delta$T$\rm_{eff}$ = 500 K and $\Delta$[Fe/H] = 0.3 dex of the solar values \citep{Soderblom_1998, Galarza_2016}. 
Solar-type stars are those similar to the Sun in mass and evolutionary stage, belonging to the F8V to K2V spectral types with 
a B-V colour in the range 0.5 to 1.0 \citep{Cayrel_1996, Soderblom_1998}. }

The stellar sample consists of 48 solar-type Galactic disk stars selected from
an updated catalogue of stellar parameters from \cite{Ramirez_2005}, with parameters
5600 K $\leq$ T$\rm_{eff}$ $\leq$  6400 K, 4.1 $\leq$ log g $\leq$ 4.6, $-$0.71 $\leq$ [Fe/H] $\leq$ $-$0.3, and 0.85 $\leq$ M $\leq$ 1.05 M$_{\odot}$. 
To have a wider metallicity range in the sample, in addition to these 48 metal-poor objects, we included 185 solar twins 
and analogues ($-$0.3 $\leq$ [Fe/H] $<$ +0.34) from our previous studies 
\citep{Spina_2018, Martos_2023, Rathsam_2023}. Thus, our final sample consists of 233 solar-type \footnote{Hereafter, we refer to the whole sample as solar-type stars because it contains all the three type stars} stars spanning the metallicity range $-$0.71 $\leq$ [Fe/H] $<$ +0.34.

For all the stars, the high-resolution spectra at R$\sim$115,000 acquired using the HARPS (High Accuracy Radial velocity Planet Searcher) spectrograph were taken from 
the European Southern Observatory (ESO) archive \footnote{\url{http://archive.eso.org/wdb/wdb/adp/phase3_spectral/form}}. 
These spectra cover the wavelength range 3780 - 6910 {\AA}. Since we need high-quality spectra
for a high-precision analysis, we selected multiple spectra of the same objects with individual S/N in the range 30 - 370. 
To minimise the effect of telluric contamination on the spectral lines, we used at least five spectra taken on
different nights. Then the spectra were radial velocity corrected using \texttt{IRAF} \footnote{Image Reduction and Analysis Facility; \url{https://iraf-community.github.io/}}, 
and then combined to get the final spectra. 
The final S/N of the combined spectra ranges from 360 to $\sim$ 1000. The combined spectra were normalised by 
splitting them into seven segments, using \texttt{IRAF}. The continuum normalisation of each individual segment was done iteratively using cubic spline 
polynomial of similar orders for the same segments, by evaluating the residuals as well as the visual fit to the continuum. 
Finally, the normalised segments are combined using \texttt{IRAF}. The spectra taken before and after 
the HARPS upgrade (June 2015) were treated separately due to the difference in the shape of the continuum, and combined afterwards. 
In addition to the spectra of the stellar sample, we also acquired HARPS solar spectra at the same resolution obtained from the asteroid Vesta.

\section{Spectroscopic analysis} \label{section_methodology}
The stellar atmospheric parameters as well as the elemental abundances were determined through a line-by-line differential 
analysis \citep{Melendez_2012, Melendez_2014, Bedell_2014} with respect to the Sun 
using the Python code \texttt{q$^{2}$} \footnote{qoyllur-quipu; \url{https://github.com/astroChasqui/q2}} \citep{Ramirez_2014}, 
adopting the line list from \cite{Melendez_2014}. The equivalent width (EW) of each spectral line is measured on 
a star-by-star basis, including those in the Sun, performing de-blending whenever necessary. For these calculations, 
\texttt{q$^{2}$} uses the ATLAS9 Kurucz grid of model atmospheres \citep{Castelli_2003} and 
the \textit{ABFIND} driver of the code MOOG (version 2019; \citealt{Sneden_1973, Sneden_2012}) 
that employs local thermodynamic equilibrium (LTE). 

\subsection{Stellar atmospheric parameters}
We used the EW measurements of 91 Fe I and 18 Fe II lines to determine the atmospheric parameters. 
The final atmospheric parameters were determined in an iterative process from the initial guess of parameters 
using the differential excitation, ionisation, and reduced EW (= EW/$\rm\lambda$) balances. The adopted solar atmospheric parameters are 
T$\rm_{eff}$ = 5777 K, log g = 4.44 dex, $\rm\zeta$ = 1 km s$^{-1}$, and [Fe/H] = 0 dex \citep{Cox_2000}. 
The code also evaluates the uncertainty on each atmospheric parameter following the procedures in \cite{Epstein_2010} and 
\cite{Bensby_2014}, which is the cumulative uncertainty that includes the uncertainty on the measurements as well as the 
uncertainty that arises due to the mutual dependence of the atmospheric parameters. The typical precisions on the estimated atmospheric 
parameters of our sample are $\sigma$(T$\rm_{eff}$) = 10 K, $\sigma$(log g) = 0.03 dex, $\sigma$($\zeta$) = 0.02 km s$^{-1}$, and 
$\sigma$([Fe/H]) = 0.007 dex. 

Following the determination of the stellar atmospheric parameters and their uncertainties, \texttt{q$^{2}$} automatically uses the appropriate 
model atmospheres to calculate the chemical abundances using the \texttt{MOOG} code under LTE.

\subsection{Chemical abundances: Mg and Y}
The abundances of the elements Mg and Y are measured for each star using the same differential EW method as above. 
All the elemental abundances are scaled relative to those obtained for the Sun on a line-by-line basis. 
We note that, in all the metal-poor stars,
the differential abundance derived from the Mg I 4730.040 {\AA} line is lower by around $\sim$ 0.05 dex compared to other Mg I lines. 
So, to get the consistent abundance values from all the Mg lines, we applied an offset of $-$4 m{\AA} to the measured EW of this line in the Sun.
The Mg I lines at 6318.717 and 6319.236 {\AA} are contaminated by nearby telluric absorption features in a few stars, and 
in those cases these lines were not considered in their abundance determination.  

Yttrium shows hyperfine splitting, and we took this into consideration when calculating its abundances. 
It is implemented by \texttt{q$^{2}$} through the \textit{blends} driver 
in the \texttt{MOOG} code and the adopted Y hyperfine lines from the Kurucz database. 
The mean errors on the estimated abundances of Mg and Y are $\rm\langle\sigma[Mg/Fe]\rangle$ = 0.01 dex and $\rm\langle\sigma[Y/Fe]\rangle$ = 0.02 dex, respectively.

\subsection{Stellar masses and ages}
The masses and ages of the stellar sample were determined using the code \texttt{q$^{2}$}, employing the 
spectroscopically determined atmospheric parameters and the Yonsei-Yale isochrones \citep{Yi_2001, Kim_2002}. 
Since our stars show non-solar abundances ([$\alpha$/Fe]$\neq$0), the [Fe/H] values do not represent their actual metallicity. 
Hence, to account for the effect of $\alpha$-enhancement on the global metallicity of the stars, 
we applied a correction factor, +log$_{10}$(0.64 $\times$ 10$^{[\alpha/Fe]}$ + 0.36) \citep{Salaris_1993}, 
together with an offset of $-$0.04 to recover the age and mass of a star with solar parameters, to the [Fe/H], 
as described in \cite{Spina_2018}. However, for the stars with [Fe/H] $<$ 0, the effect of $\alpha$-enhancement is 
already incorporated in the Yonsei-Yale isochrones to follow the Galactic trend, using the [$\alpha$/Fe] values given in 
\cite{Melendez_2010}. We had to subtract this factor to avoid double $\alpha$-enhancement. 
For the stars in the metallicity range $-$1 $\leq$ [Fe/H] $\leq$ 0, the value is given by [$\alpha$/Fe] = $-$0.3 $\times$ [Fe/H]. 
So, the final global metallicity ([M/H]) of the stars in our sample is given by Eq. \ref{equation_global_metallicity}, where we have used 
our estimates of [Mg/Fe] as a proxy for [$\alpha$/Fe]:

\begin{equation}
\label{equation_global_metallicity} 
\begin{split}
     \rm [M/H] = \rm [Fe/H] &- 0.04 + \rm log_{10} (0.64 \times 10^{\rm [Mg/Fe]} + 0.36) \\ 
           &- \rm log_{10} (0.64 \times 10^{-0.3 \times \rm [Fe/H]} + 0.36). \\
\end{split}
\end{equation}

The \texttt{q$^{2}$} code produces a probability distribution for stellar masses and ages through an isochrone fitting, 
where the spectroscopic stellar parameters are compared with those obtained from the grid of isochrones \citep{Ramirez_2014}.  
The most probable values of these probability distributions are adopted as stellar masses and ages, and 
the 1-$\sigma$ range around the most probable values are taken as their respective uncertainties.

For this calculation, we used the parallax values adopted from the \textit{Gaia} DR3 \footnote{\url{https://gea.esac.esa.int/archive/}} \citep{Gaia_2016b, Gaia_2022K, Babusiaux_2022}, 
and the V magnitudes from SIMBAD. If the V magnitudes were not available, then we estimated it using the \textit{Gaia} G magnitudes and (BP - RP) colour indices 
through the calibration equation given by \cite{Rathsam_2023}. 

The ages and masses of the solar twins from \cite{Spina_2018} were recalculated using the same procedure as our sample, for consistency in the analysis, 
and are listed in \cite{Martos_2023}. {The typical uncertainties in the isochrone ages and masses of the stellar sample are $\sim$0.4 Gyr and 0.01 M$_{\odot}$, respectively.}

\section{Results and discussion} \label{section_results_discussion}
Here, we discuss the results of our analysis of the combined sample of 233 
disk stars (48 solar-type stars from this study and 185 solar twins and analogues from our previous studies). 
The estimated stellar atmospheric parameters, masses, ages, and the abundance ratios of our sample are 
given in Table \ref{Table_results_1}, and those of solar twins and analogues are given in Table \ref{Table_results_2}. 
The parameters of the combined sample are 5149 K $\leq$ T$\rm_{eff}$ $\leq$ 6210 K, 4.04 $\leq$ log g $\leq$ 4.57, 
$-$0.71 $\leq$ [Fe/H] $<$ 0.34, and 0.82 $\leq$ M $\leq$ 1.1 M$_{\odot}$, and all the stars lie in the 
solar neighbourhood (R$_{GC}$ $\sim$ 8 kpc). 

\subsection{Bi-modality of the Galactic disk and age--metallicity relation}

\begin{figure}
\centering
\includegraphics[width=\columnwidth]{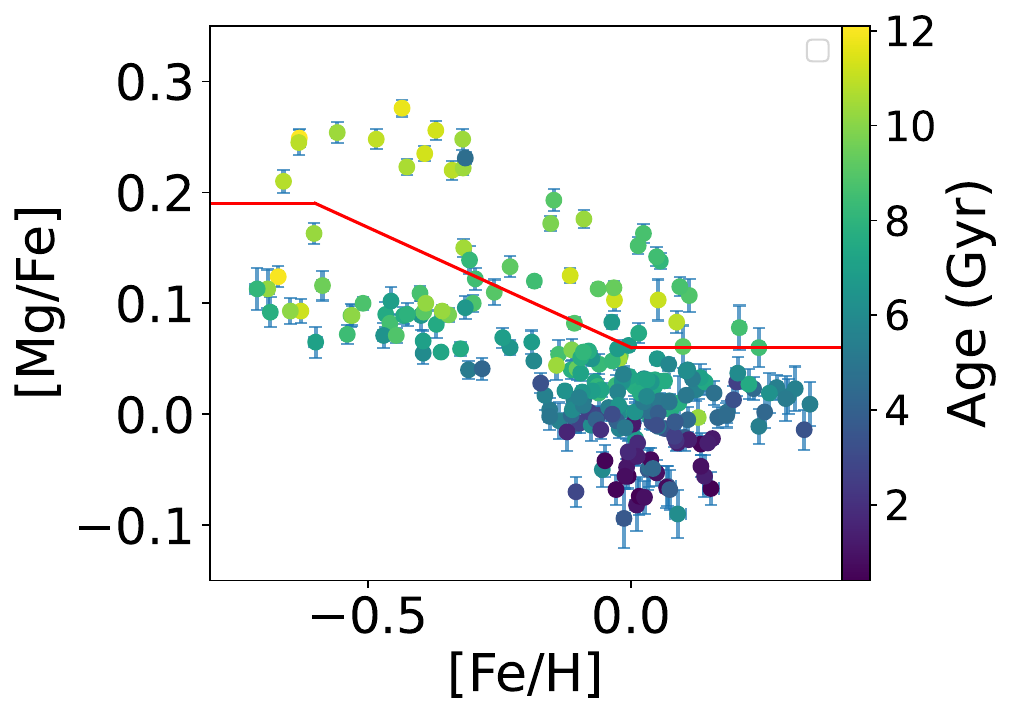} 
\includegraphics[width=\columnwidth]{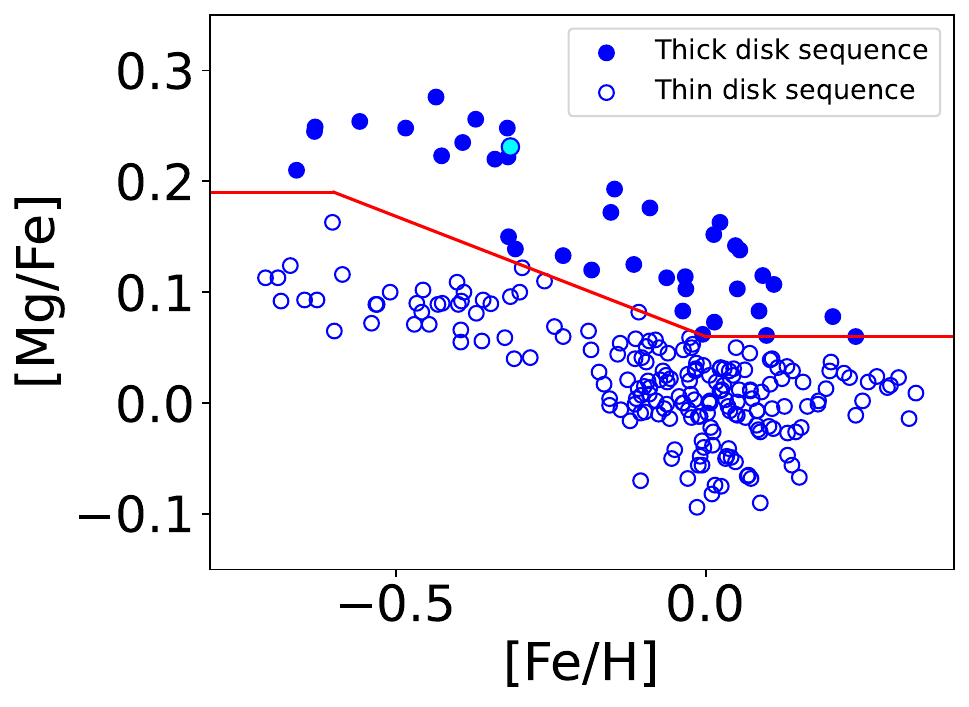} 
\caption{[Mg/Fe] ratio as a function of [Fe/H] for the combined sample. 
The solid red line shows the separation between thin and thick disk stars based on the [Mg/Fe] ratio, similar to \cite{Adibekyan_2011}:
[Mg/Fe] = +0.19 for [Fe/H] $\leq$ $-$0.6, [Mg/Fe] = $-$0.217 $\times$ [Fe/H] + 0.06 for $-$0.6 $<$ [Fe/H] $\leq$ 0, and [Mg/Fe] = +0.06 for [Fe/H] $>$ 0.   
Stars above this line belong to the Galactic thick disk. The cyan circle is the anomalous star HD 65907.}\label{Thin_Thick_disk_limit}
\end{figure}

Figure \ref{Thin_Thick_disk_limit} shows the [Mg/Fe] ratio with respect to the [Fe/H] of the combined sample. 
From this figure, we can clearly see two populations of [Mg/Fe] at low-metallicity, the high-$\alpha$ or thick disk 
and the low-$\alpha$ or thin disk. At higher metallicities ([Fe/H] $\geq$ 0), both populations are indistinguishable. 
This bimodality of the Galactic disk in the [$\alpha$/Fe] - [Fe/H] plane has already been identified in several previous 
studies (e.g. \citealt{Reddy_2006, Adibekyan_2011, Adibekyan_2012, Haywood_2013, Nivedar_2014, Bensby_2014, Ness_2019, Weinberg_2019, Xiang_2022, Leung_2023, Imig_2023, Patil_2023}). 
It is the consequence of different chemo-dynamical evolution and star formation history of the Milky Way disk (e.g. \citealt{Haywood_2013, Bergemann_2014, Casagrande_2011}).
{Generally, thick disk is composed of old, metal-poor, $\alpha$-enhanced stars, whereas thin disk is composed of 
younger, $\alpha$-poor stars \citep{Fuhrmann_1998, Bensby_2005, Lee_2011}. However, there is no obvious method for completely disentangling these two components of the Galactic disk and 
to identify purely thin or thick disk stars in the solar neighbourhood \citep{Bensby_2003, Adibekyan_2012, Haywood_2013}. 
Furthermore, the separation between the $\alpha$-sequences and even the existence of boundary between these two populations \citep{Bovy_2012a} are 
still topics of debate. A detailed review on Galactic disk can be found in \cite{Rix_2013}. }

{Previously, the thin and thick disk stars were distinguished either from a pure kinematical approach (e.g. \citealt{Bensby_2003, Reddy_2006}) 
or from an approach that combines kinematics, metallicity, and age (e.g. \citealt{Haywood_2008}). However, both these methods are found to be prone to
$\textquoteleft$contamination' or overlap between the two components \citep{Bensby_2003, Adibekyan_2012, Bovy_2012b}.
In addition to this, the radial migration is proposed to have redistributed the stars in the solar neighbourhood (e.g. \citealt{Liu_2012, Haywood_2013}).
This radial migration and the higher eccentric orbits lead to the contamination of 
the kinematically selected thin and thick disk stars in the solar vicinity (e.g. \citealt{Schonrich_2009}). }

{The advent of several large sky multi-object spectroscopic surveys in the last decade allowed the chemo-kinematic 
analysis of much larger samples of stars. As a consequence, a purely chemical approach to dissecting the Galactic disk began to gain significance. 
Several studies have shown that chemical dissection of the Galactic disk is more reliable 
because the chemical composition of a star is a more stable property that is closely related to the 
time and place of birth, whereas the kinematics and spatial positions change all 
the time \citep{Navarro_2011, Adibekyan_2011, Liu_2012, Bovy_2012b, Adibekyan_2012, Adibekyen_2013}.
It has also been proposed that various properties (scale height, kinematics, etc) of different Galactic stellar populations 
can be parametrised using their chemical composition \citep{Bovy_2012b}. }

{For a given age, the solar vicinity exhibits a broader metallicity distribution since stars of different 
birth radii (and hence different metallicities) contribute to the local neighbourhood \citep{Liu_2012}. 
As a result, despite the fact that the thick disk is comparatively older, stars with a range of ages and metallicities 
can contribute to both thin and thick disks, giving rise to a large overlap in metallicity.
Hence, even though the chemical composition is the more reliable approach, the metallicity ([Fe/H]) alone is not an ideal choice. 
From the chemo-orbital analysis of a sample of SDSS/SEGUE data (Sloan Extension for Galactic Understanding and Exploration; \citealt{Yanny_2009}), \cite{Liu_2012} showed that the $\alpha$-abundance 
is preferred because it is relatively independent of birth radii and is a more appropriate proxy for age. The 
distribution of the [$\alpha$/Fe] ratio with respect to [Fe/H] has been used in several studies to distinguish between the two 
disk populations (e.g. \citealt{Adibekyan_2011, Haywood_2013}). 
The limit characterising the thin and thick disk populations is defined empirically, 
by examining the chemical separation in [$\alpha$/Fe] between the high-$\alpha$ (thick disk) and low-$\alpha$ (thin disk) sequences.
Recent studies based on the chemo-kinematical and 
statistical methods using large samples from high-resolution spectroscopic surveys, such as Apache Point Observatory Galactic Evolution Experiment (APOGEE; e.g. \citealt{Anders_2014, Patil_2023}), the 
\textit{Gaia}-ESO Survey (GES; e.g. \citealt{Gent_2024}), and GALactic Archaeology with HERMES (GALAH; e.g. \citealt{Cantelli_2024}), have shown that the decomposition of the Galactic disk is 
better understood in terms of the chemical enrichment ($\alpha$ abundance).}

{On the grounds of the above discussion, we adopted the chemical separation given in \cite{Adibekyan_2011} to separate the thin and thick disk sequences. 
This is shown as a solid red line in the [Mg/Fe] - [Fe/H] plane in Fig. \ref{Thin_Thick_disk_limit}. We identified the stars above 
this line as thick disk stars and those below as thin disk stars \footnote{Chemically defined}. 
The behaviour of [Mg/Fe] ratio with respect to various orbital properties and the Toomre diagram of our sample are 
shown in Figs. \ref{MgFe_orbital_parameters} and \ref{Toomre_diagram}. As seen in the several previously mentioned works, 
our chemically defined thick and thin disk stars are mixed in the kinematic plane.
The thick disk contains the older stars (age $>$ 8 Gyr; two stars have ages of 6.7 and 7.5 Gyr) of the sample,} whereas the thin disk shows an age spread. 
All the youngest stars of the sample belong to the thin disk.

We note that a younger star with an age of 4.6 Gyr is a member of the thick disk (cyan circle; Fig. \ref{Thin_Thick_disk_limit} lower panel). 
This object, HD 65907, shows anomalous behaviour compared to other stars of the sample. As seen in the figures, 
it lies well off the other objects of its kind. 
\cite{Fuhrmann_2012} analysed this anomalous object, 
and from its chemical abundance they identified it as an old, population II, thick disk star with an age of 12 or 13 Gyr. 
However, the age inferred from the evolutionary tracks is 5.6$^{+2.2}_{-1.8}$ Gyr, which contradicts its chemical properties. 
According to \cite{Fuhrmann_2012}, this discrepancy can only be resolved with a mass accretion scenario on HD 65907.  
So, this object is probably a product of a former mass-transfer event, like the solar analogue HIP 10725 \citep{Schirbel_2015}.

\begin{figure}
\centering
\includegraphics[width=\columnwidth]{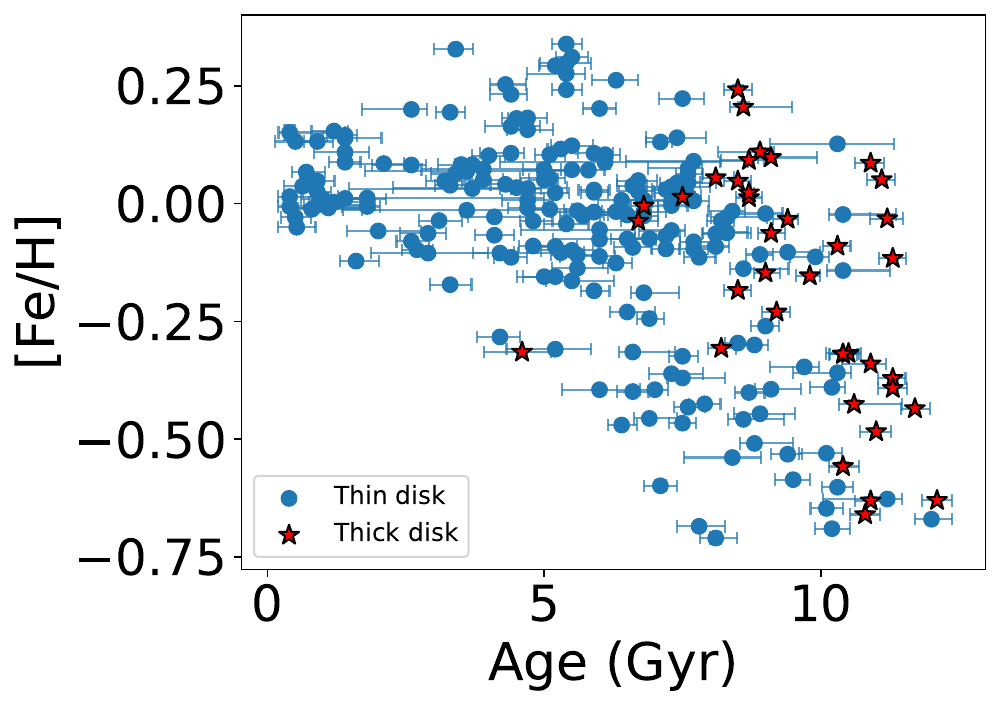} 
\includegraphics[width=\columnwidth]{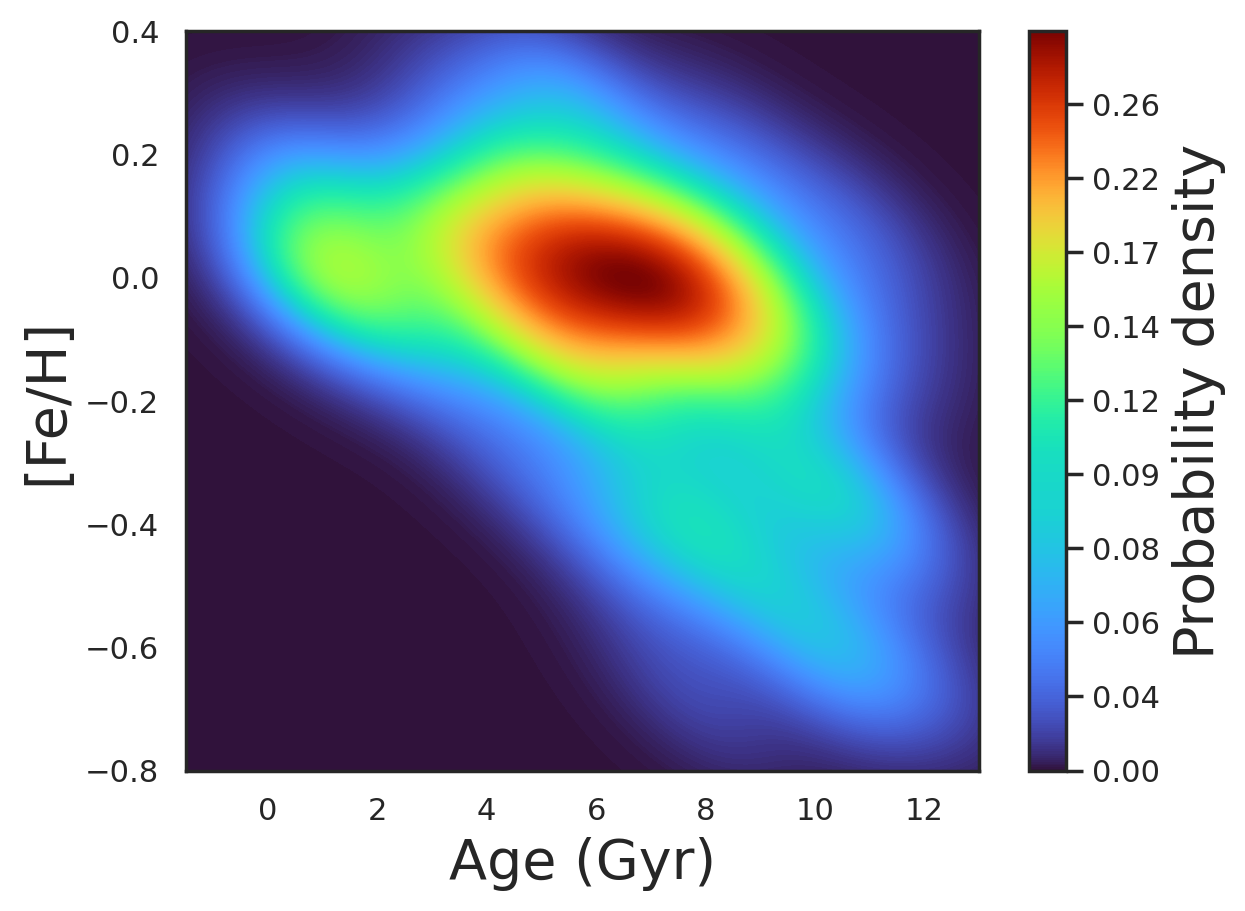} 
\caption{AMD of the sample.
The red star that is far from the other thick disk objects in the upper panel is the anomalous star HD 65907.
The bottom panel is the density plot for the AMD of the sample. 
}\label{AMR_Thin_Thick_disk}
\end{figure}

The age--metallicity distribution (AMD) of the whole sample is shown in Fig. \ref{AMR_Thin_Thick_disk}. 
Earlier studies found that the age and metallicity are not strongly correlated in 
the solar neighbourhood, and that the metallicity shows a significant scatter for the same 
age (e.g. \citealt{Edvardsson_1993, Feltzing_2001, Haywood_2006, Casagrande_2011, Bensby_2014, Lin_2020, Sahlholdt_2022, Patil_2023}).
From Fig. \ref{AMR_Thin_Thick_disk}, it is clear that stars of similar age show a large range of metallicities. 
The thick disk stars in the sample show an age--metallicity relation (AMR; Fig. \ref{AMR_Thin_Thick_disk}, upper panel), as previously noted 
by \cite{Schuster_2006}, \cite{Bensby_2014}, \cite{Patil_2023}, and others. Meanwhile, the age and metallicity are uncorrelated for the
thin disk stars, and show an almost flat distribution. However, both the populations are indistinguishable
from each other in the age--[Fe/H] space. 

From the contour plot for the AMD of the stellar sample (bottom panel, Fig. \ref{AMR_Thin_Thick_disk}), 
we note the following: (i) up to the age of $\sim$ 12 Gyr and for [Fe/H] $>$ $-$0.2, we see a flat AMR with a scatter in the metallicity;
(ii) for the older (age $>$ 7 Gyr), metal-poorer stars in the sample ([Fe/H] $<$ $-$0.2), we see a downward trend in the AMD; and
(iii) the majority of the distribution is centred at [Fe/H] $\sim$ 0 with a peak around ages between 4 - 9 Gyr. 
This is on par with the findings in the previous studies of the Galactic disk (e.g. \citealt{Edvardsson_1993, Feltzing_2001, Bergemann_2014}).
The AMR and the evolution of the Galactic disk are broader topics, and their detailed discussion is beyond the scope of this paper. 
Several studies with large samples have been  dedicated to the AMR in the Galactic disk, including \cite{Feltzing_2001}, \cite{Lin_2020}, \cite{Sahlholdt_2022}, and \cite{Patil_2023}, to name a few.

\subsection{Abundance trend with age}
\begin{figure}
\centering
\includegraphics[width=\columnwidth]{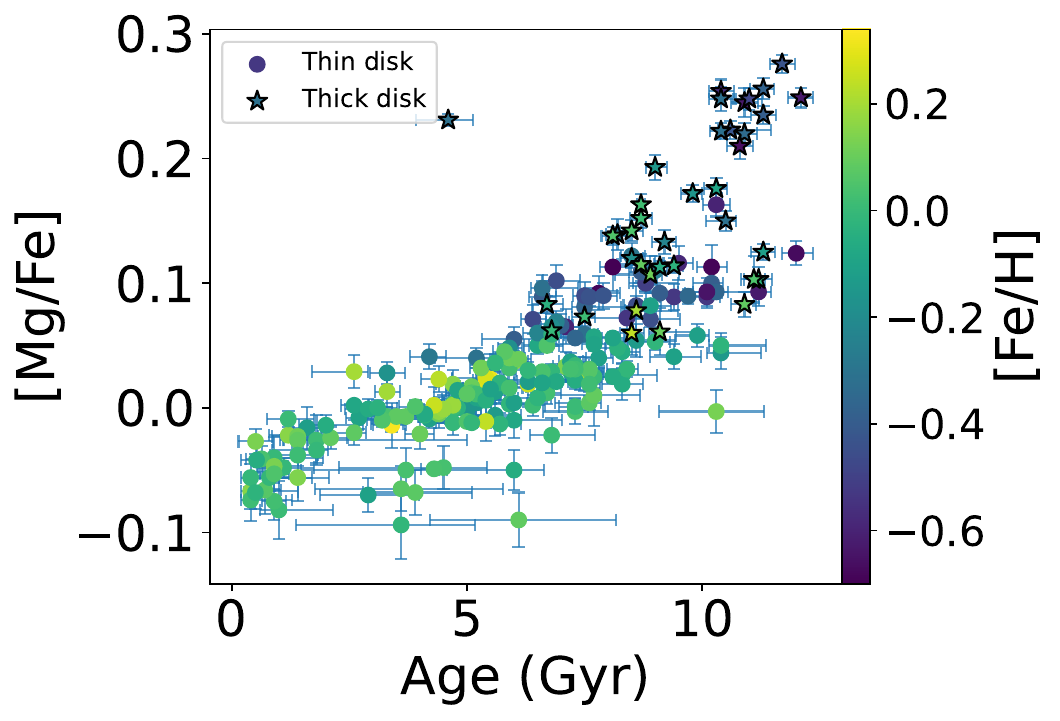} 
\includegraphics[width=\columnwidth]{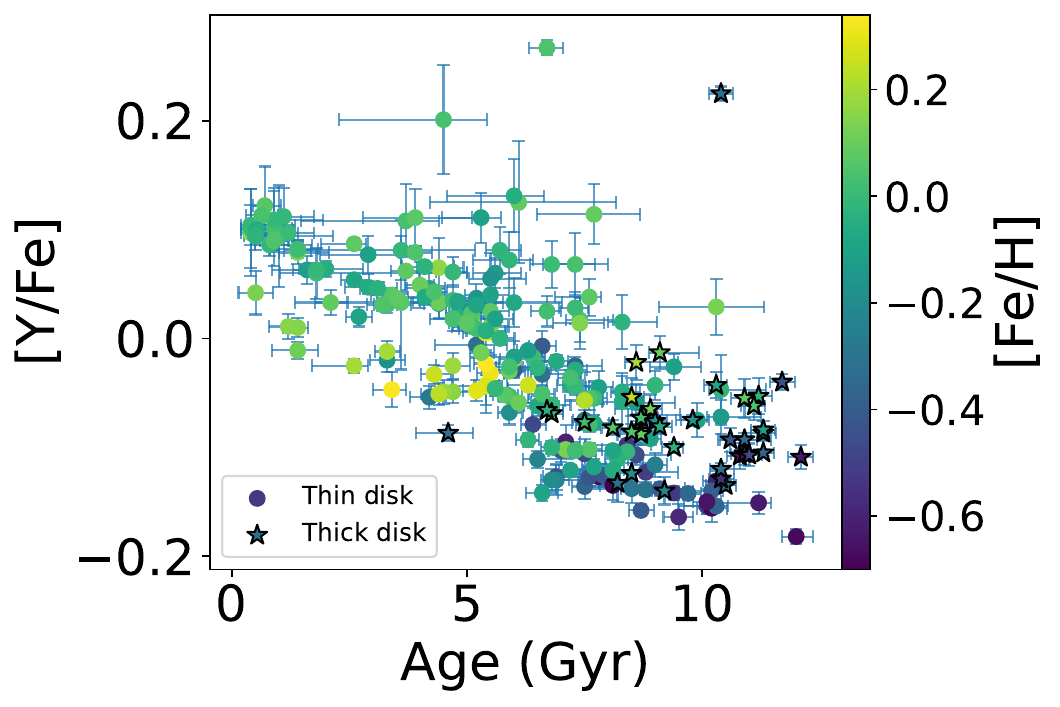}
\includegraphics[width=\columnwidth]{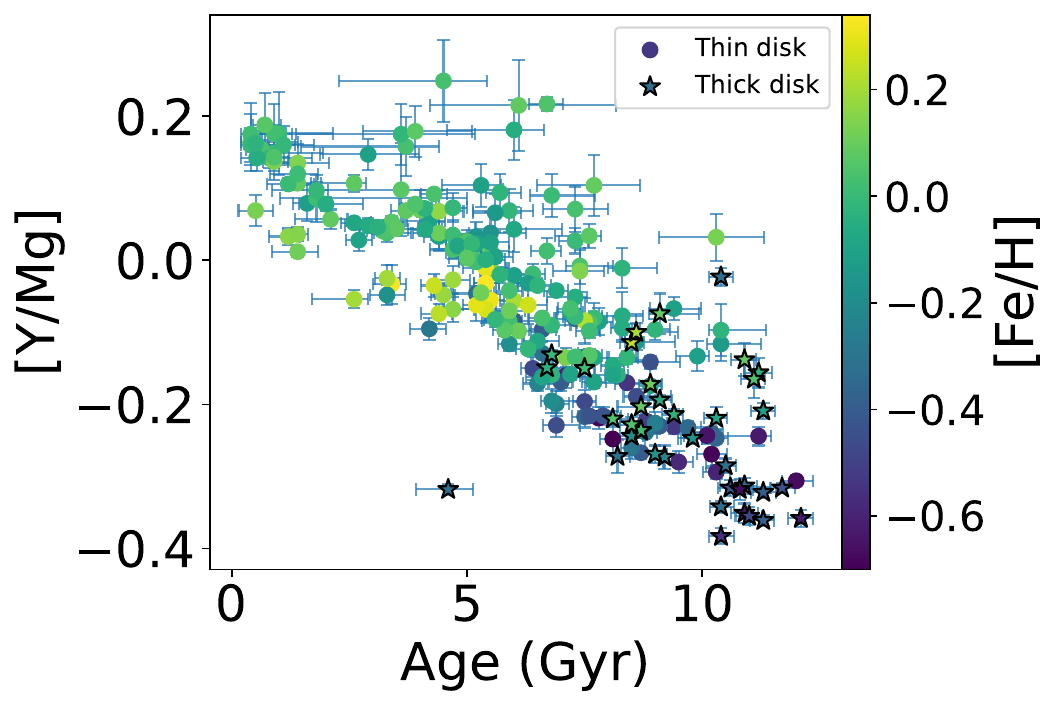} 
\caption{[Mg/Fe], [Y/Fe], and [Y/Mg] ratios as a function of age for the combined sample, colour-coded by [Fe/H]. 
The star at $\sim$ 5 Gyr is HD 65907 and is probably the result of a former mass-transfer event.}\label{YFe_MgFe_all_sample}
\end{figure}

The distributions of [Mg/Fe], [Y/Fe], and [Y/Mg] with respect to stellar ages of the combined sample are shown in 
Fig. \ref{YFe_MgFe_all_sample}. As we see from the [Mg/Fe]--[Fe/H] distribution (Fig. \ref{Thin_Thick_disk_limit}), the thin and thick disk objects 
show a clear dichotomy in the [Mg/Fe]--age plane as well. 
The [Mg/Fe] and [Y/Fe]  ratios show positive and negative slopes, respectively, with respect to age, as expected from
their origin and chemical evolution models. Mg is an $\alpha$-element produced by {Type II supernovae (SNe II)} on a shorter timescale ($\leq$ 10$^{7}$ yrs; e.g. \citealt{Kobayashi_2006}). 
SNe II were the main contributors of metals to the {interstellar medium} during the early stages of the chemical evolution. So, $\alpha$-elements show a positive trend with respect to age. 
Y is an s-process element produced in low- and intermediate-mass {asymptotic giant branch} stars, which contributed to the chemical evolution much later compared to SNe II. 
As a result, Y shows a negative trend with stellar ages (\citealt{Nomoto_2013, Matteucci_2014, Karakas_2014, Spina_2016b} and references therein).

As a consequence of different origins and chemical evolution, the [Y/Mg] ratio has a steeper dependence on stellar ages. 
As seen from Fig. \ref{YFe_MgFe_all_sample} (bottom panel), the [Y/Mg] ratio of our sample shows a strong anti-correlation 
with age. The scatter in the [Y/Mg] ratio of our sample is small compared to that found in other samples of 
solar-type stars in the literature (e.g. \citealt{Feltzing_2017, Titarenko_2019, Casali_2020}). 
We note continuity in the [Y/Mg]--age correlation between the thin and thick disk stars. 
We also note that there is no significant gradient of [Y/Mg] with respect to [Fe/H], as was seen in \cite{Feltzing_2017}.

\subsection{The [Y/Mg] chemical clock}
\begin{figure}
\centering
\includegraphics[width=\columnwidth]{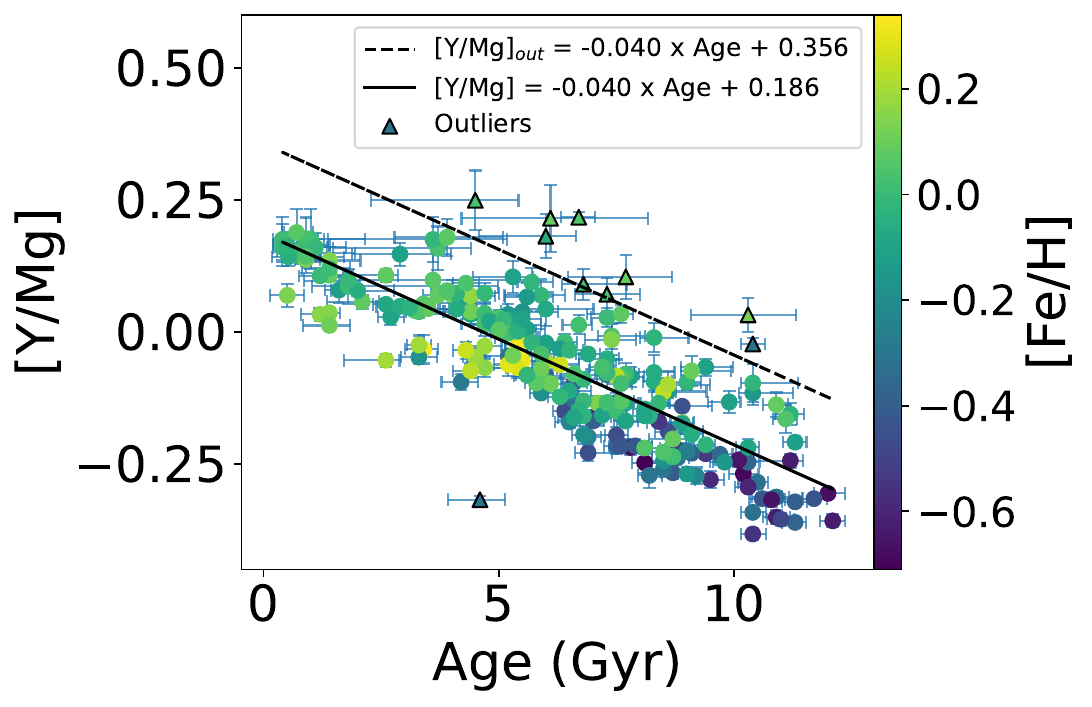} 
\caption{Linear fit to the [Y/Mg]--age distribution of the whole sample (solid black line) and the outlier cut (dashed black line). 
The outlier limit is set at the 2$\sigma$ level of the trend. The triangle in the lower part of the plot is the anomalous star HD 65907.}\label{YMg_outlier_limit}
\end{figure}

\begin{figure}
\centering
\includegraphics[width=\columnwidth]{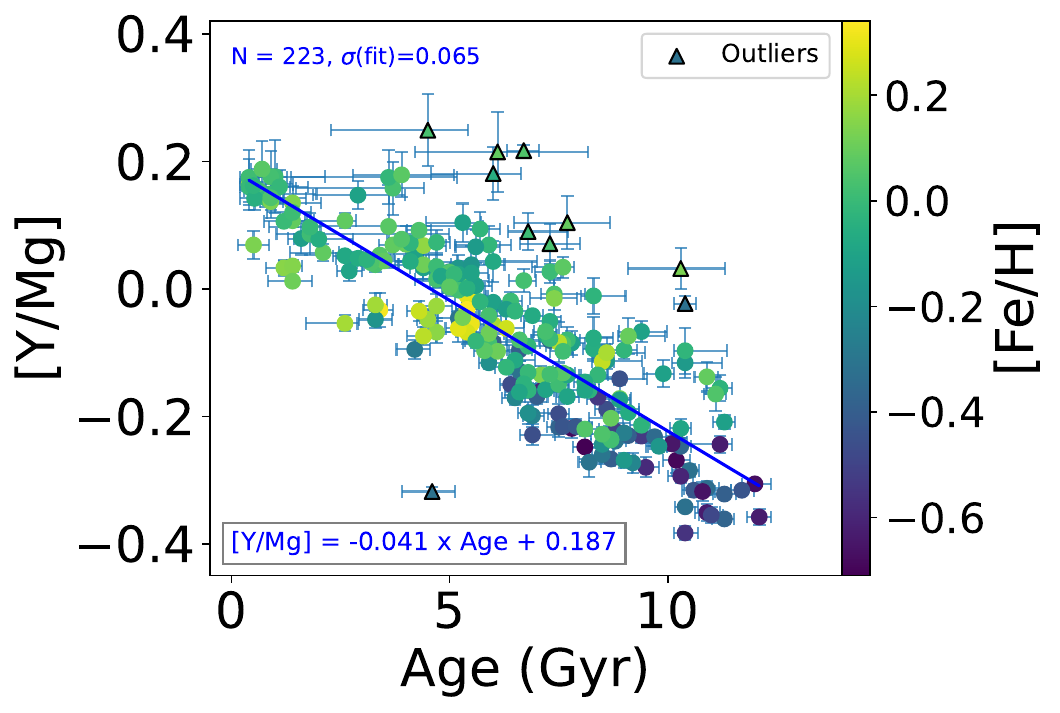} 
\caption{Best linear fit (solid blue line) for the [Y/Mg]--age distribution of the sample. The fitting function is described in the text (Eq. \ref{equation_chemical_clock}).}\label{YMg_linear_fit}
\end{figure}

In this section we analyse the [Y/Mg]--age correlation in our sample. The stellar ages derived from isochrone fitting 
were used in this analysis. We performed a simple linear regression fitting to derive the empirical relation connecting the [Y/Mg] ratio and 
stellar ages, of the form [Y/Mg] = a $\times$ Age + b. {From the linear fitting, we find that the [Y/Mg] of the sample is correlated to the stellar age through the 
relation (Fig. \ref{YMg_outlier_limit}) 
\begin{equation}
\label{equation_chemical_clock_whole} 
\begin{split}
   \rm [Y/Mg] = -0.040(\pm0.001) \times \rm Age \,(Gyr) + 0.186(\pm0.036). \\
\end{split}
\end{equation} 
}To get the best possible relation, we eliminated a few stars that lie outside the bulk of the sample 
in the [Y/Mg]--age space. {The outlier limit is set at the 2$\sigma$ level of the trend (Eq. \ref{equation_chemical_clock_whole}), which is given by the linear function [Y/Mg] = $-$0.040 $\times$ Age (Gyr) + 0.356, as shown in Fig. \ref{YMg_outlier_limit}.} We also excluded the anomalous star
HD 65907 from this analysis. We did not make any distinction between the thin and thick disk objects. From the linear fit 
shown in Fig. \ref{YMg_linear_fit} for the sample after excluding the outliers, we obtained a relation of the form
\begin{equation}
\label{equation_chemical_clock} 
\begin{split}
   \rm [Y/Mg] = -0.041(\pm0.001) \times \rm Age \,(Gyr) + 0.187(\pm0.040). \\
\end{split}
\end{equation}

\begin{figure}
\centering
\includegraphics[width=\columnwidth]{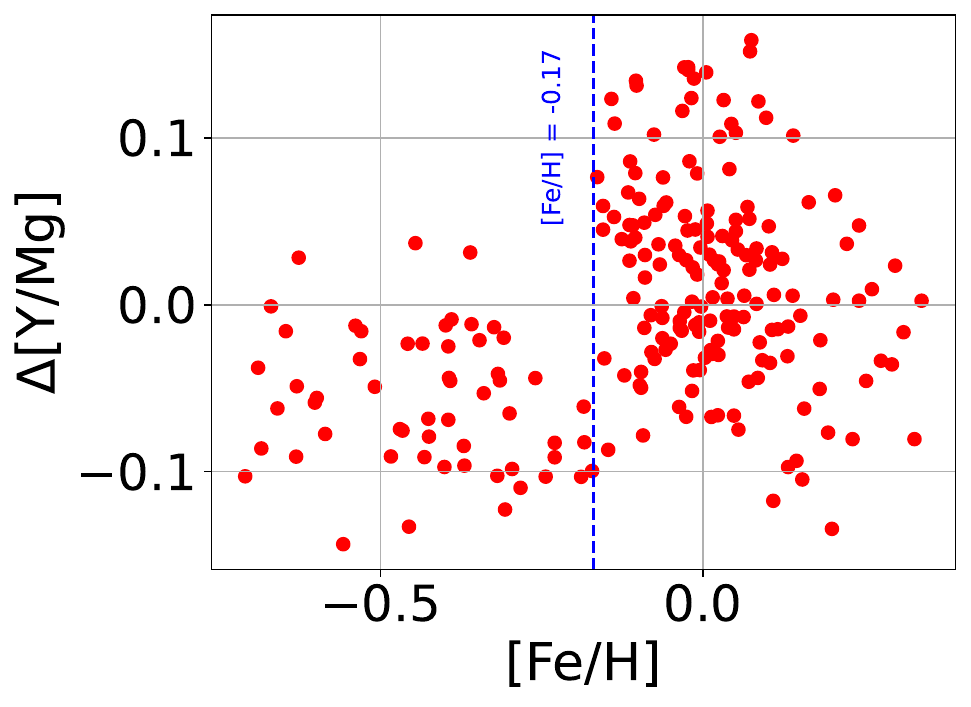} 
\caption{Distribution of the [Y/Mg] residuals ($\rm\Delta[Y/Mg]$) with respect to the metallicity of the sample. 
The vertical dashed line at [Fe/H] = $-$0.17 represents the point below which $\rm\Delta[Y/Mg]$ shows an offset with metallicity.}\label{ymg_residual}
\end{figure}

\begin{figure}
\centering
\includegraphics[width=\columnwidth]{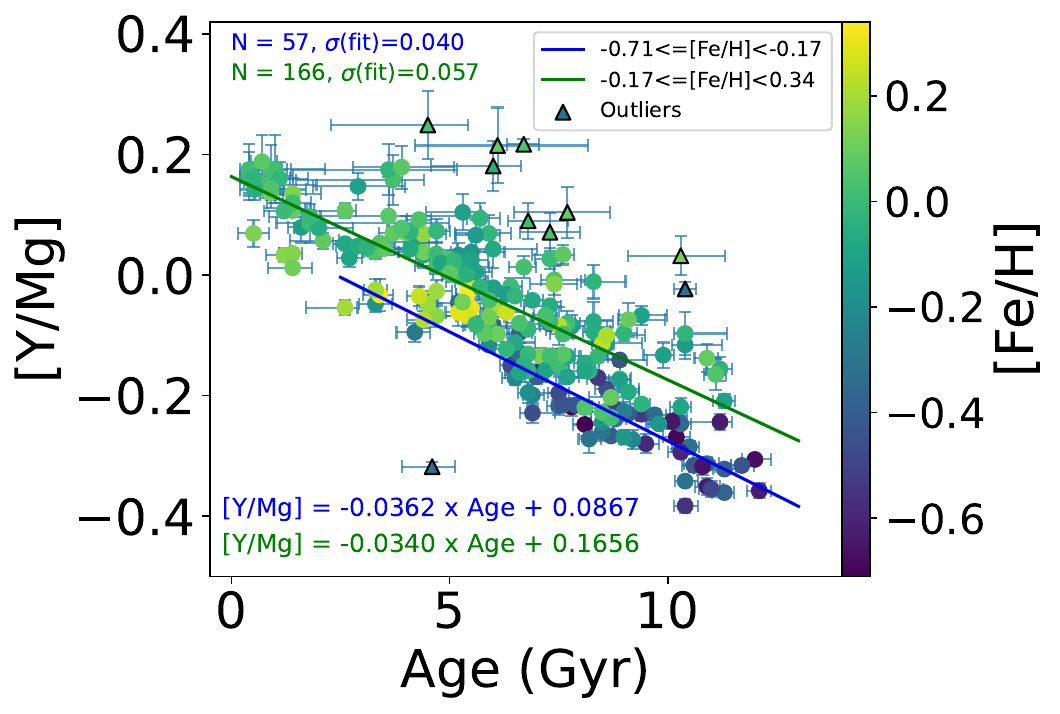} 
\caption{[Y/Mg]--age linear fits for the sample in two metallicity bins: $-$0.71 $\leq$ [Fe/H] $<$ $-$0.17 (solid blue line) and $-$0.17 $\leq$ [Fe/H] $<$ +0.34 (solid green line).}\label{YMg_feh_bins}
\end{figure}

However, from the analysis of a sample of 714 solar-type stars, \cite{Feltzing_2017} first identified the metallicity dependence of 
the [Y/Mg] chemical clock, which was subsequently found in the samples of \cite{Delgadomena_2019} and  \cite{Casali_2020}. 
To verify the metallicity dependence of Eq. \ref{equation_chemical_clock}, we examined the distribution of 
[Y/Mg] residuals ($\rm\Delta[Y/Mg]$ = [Y/Mg]$_{obs}$ - [Y/Mg]$_{pred}$) with respect to metallicity to see if there exist any trend. 
As we can see from Fig. \ref{ymg_residual}, the $\rm\Delta[Y/Mg]$ is independent of metallicity up to a value of [Fe/H] $\sim$ $-$0.17 and show a trend below that. 
This confirms that the [Y/Mg] chemical clock depends on the metallicity and that the whole sample cannot be represented using the single relation given in Eq. \ref{equation_chemical_clock}. 
Hence, we divided the sample into two metallicity bins, $-$0.71 $\leq$ [Fe/H] $<$ $-$0.17 and $-$0.17 $\leq$ [Fe/H] $<$ +0.34, and performed 
the fitting separately, as shown in Fig. \ref{YMg_feh_bins}. This simple linear fitting resulted in the following relations: 
\begin{equation}
\label{equation_chemical_clock_feh_bins} 
\begin{split}
   \rm [Y/Mg] = -0.0362(\pm0.0004) \times \rm Age \,(Gyr) + 0.0867(\pm0.0285);  \\ (\rm for -0.71 \leq [Fe/H] < -0.17) \\
    \rm [Y/Mg] = -0.0340(\pm0.0006) \times \rm Age \,(Gyr) + 0.1656(\pm0.0346); \\ (\rm for -0.17 \leq [Fe/H] < +0.34). \\
\end{split}
\end{equation}
We performed this fitting for five different metallicity bins of width 0.2 dex as well, and note that the results  
do not produce significant deviations from the results obtained using Eq. \ref{equation_chemical_clock_feh_bins}.

We then inverted the relations in Eq. \ref{equation_chemical_clock_feh_bins} to get the stellar dating relations of the form
\begin{equation}
\label{equation_dating_relation} 
\begin{split}
   \rm Age \, (Gyr) = -27.6243(\pm0.3053) \times \rm [Y/Mg] + 2.3950(\pm0.8139); \\ (\rm for -0.71 \leq [Fe/H] < -0.17) \\
    \rm Age \, (Gyr) = -29.4118(\pm0.5191) \times \rm [Y/Mg] + 4.8706(\pm1.1039); \\ (\rm for -0.17 \leq [Fe/H] < +0.34). \\ 
\end{split}
\end{equation}

\begin{figure}
\centering
\includegraphics[width=\columnwidth]{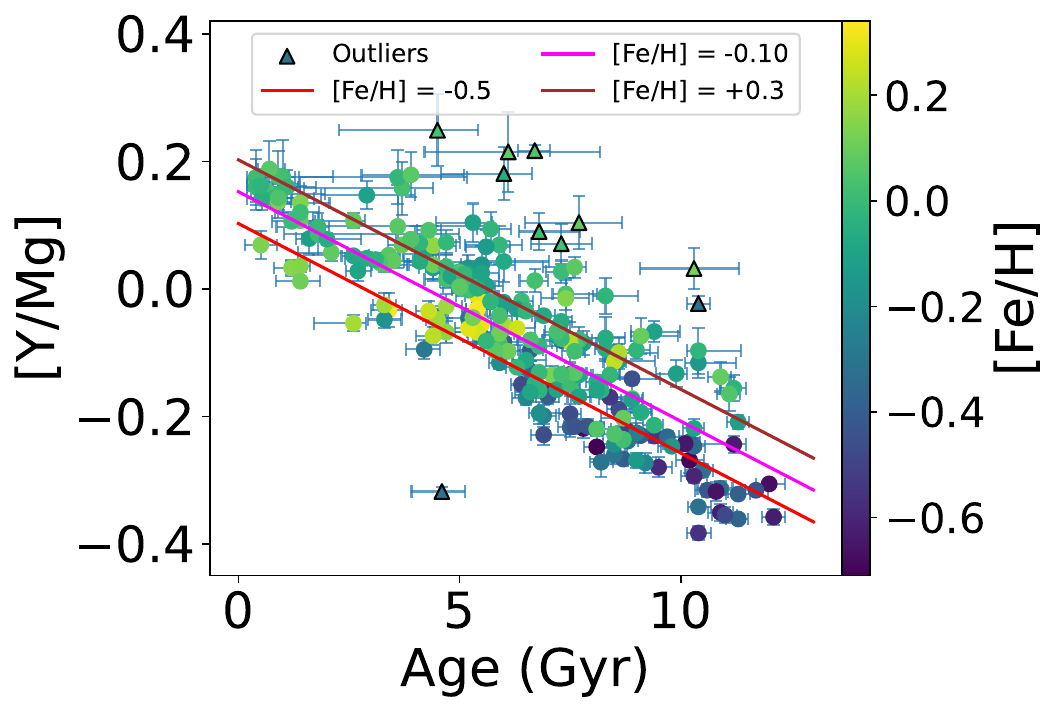} 
\caption{MV linear fits for the [Y/Mg]--age distribution of the sample for three different metallicities.  
The fitting function is described in the text (Eq. \ref{equation_chemical_clock_multivariate}).}\label{YMg_multivariate_linear_fit}
\end{figure}

{Since we identified the metallicity dependence of the [Y/Mg] ratio in the sample, we also performed a multivariate (MV) linear regression fitting by 
incorporating the [Fe/H] as well. The obtained relation as a function of stellar age and 
metallicity is given by 
\begin{equation}
\label{equation_chemical_clock_multivariate} 
\begin{split}
   \rm [Y/Mg] =  & -0.036(\pm0.001) \times \rm Age \,(Gyr)  \\ 
                &+ 0.125(\pm0.010) \times \rm [Fe/H] + 0.165(\pm0.034). \\
\end{split}
\end{equation} 
The chemical age is given by the relation 
\begin{equation}
\label{equation_dating_relation_multivariate} 
\begin{split}
    \rm Age \, (Gyr) = & -27.701(\pm0.461) \times \rm [Y/Mg] \\ 
                        &+ 3.454(\pm0.329) \times \rm [Fe/H] + 4.576(\pm1.013). \\ 
\end{split}
\end{equation}
Figure \ref{YMg_multivariate_linear_fit} shows the relation in Eq. \ref{equation_chemical_clock_multivariate} for three different metallicity values, [Fe/H] = $-$0.5, $-$0.1, and +0.3. }

\begin{figure}
\centering
\includegraphics[width=\columnwidth]{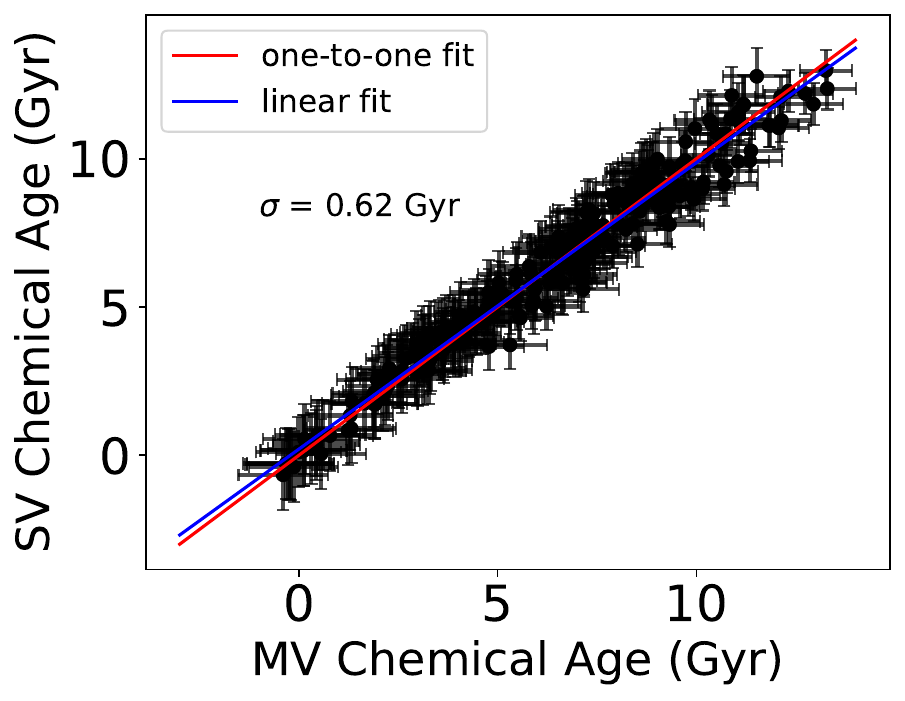} 
\caption{Correlation between the SV (Eq. \ref{equation_dating_relation}) and 
MV (Eq. \ref{equation_dating_relation_multivariate}) chemical ages of the stellar sample.}\label{SV_MV_age}
\end{figure}

{The chemical ages of the sample are calculated using both Eq. \ref{equation_dating_relation} and Eq. \ref{equation_dating_relation_multivariate}
and compared with the isochrone ages and with each other. The chemical ages do not show any trend with the isochrone ages (offset $\sim$ 0). 
The standard deviation (SD) of the 
single variate (SV) chemical age (Eq. \ref{equation_dating_relation}) with respect to the isochrone age is 1.55 Gyr and that of the 
MV chemical age (Eq. \ref{equation_dating_relation_multivariate}) is 1.66 Gyr. We note that the average uncertainty in 
the isochrone ages of the sample is $\sim$0.4 Gyr. From the comparison between
the SV and MV chemical ages shown in Fig. \ref{SV_MV_age},  we see that the two ages match quite well, with an offset of only $-$0.005 Gyr. The mean scatter and SD between 
them are 0.50 and 0.62 Gyr, respectively. }

\subsubsection{Validation using open clusters and stars with asteroseismic ages}  
{To validate our relations in Eqs. \ref{equation_dating_relation} and \ref{equation_dating_relation_multivariate}, we used the 
data for open clusters (OCs) and stars with asteroseismic ages. OCs are one of the powerful tools for validating the chemical clocks because the member stars of each 
cluster show homogeneity in age and chemical composition. As a result, their ages have been precisely determined through isochrone fitting from the observation 
of several member stars (e.g. \citealt{Casali_2019, Viscasillas_2022}). \cite{Casali_2020} present a list of OCs selected from the iDR5 (fifth internal data release) of the GES (\citealt{Gilmore_2012, Randich_2013}), with cluster membership probability of at least 80\%. 
From this list, we selected 13 OCs with R$_{GC}$ $\geq$ 6.5 kpc.}

{In the case of single stars, asteroseismology provides the reliable ages from the solar-like oscillations 
on the stellar surface \citep{Chaplin_2013, Silva_2012, Christensen_2016}.  \cite{Silva_2017} present the asteroseismic ages for a 
sample of 66 main-sequence stars from the \textit{Kepler} LEGACY sample (KLS) based on the asteroseismic data from the \textit{Kepler} mission.
The KLS is claimed to have the best asteroseismic data available for solar-like stars. We selected ten stars from the KLS for which 
the precise abundances are available in the literature.
The validation sample of OCs and KLS stars used to calibrate our chemical clocks are given in Table  \ref{Table_OC_KLS}. 
The Y and Mg abundances of OCs and KLS stars are adopted from \cite{Casali_2020} and \cite{Nissen_2017}, respectively. }

\begin{figure}
\centering
\includegraphics[width=\columnwidth]{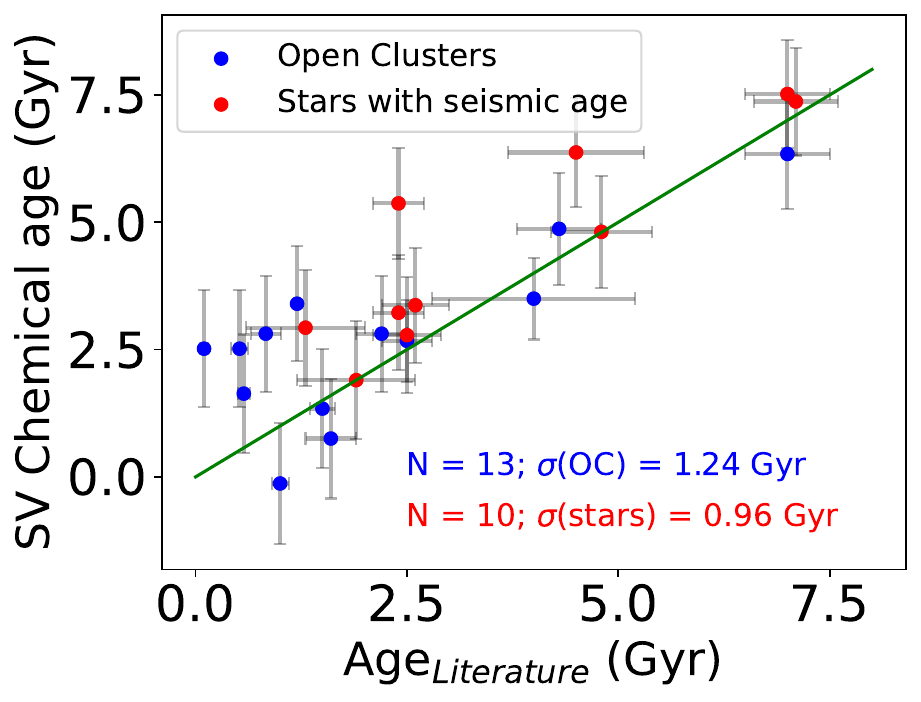} 
\includegraphics[width=\columnwidth]{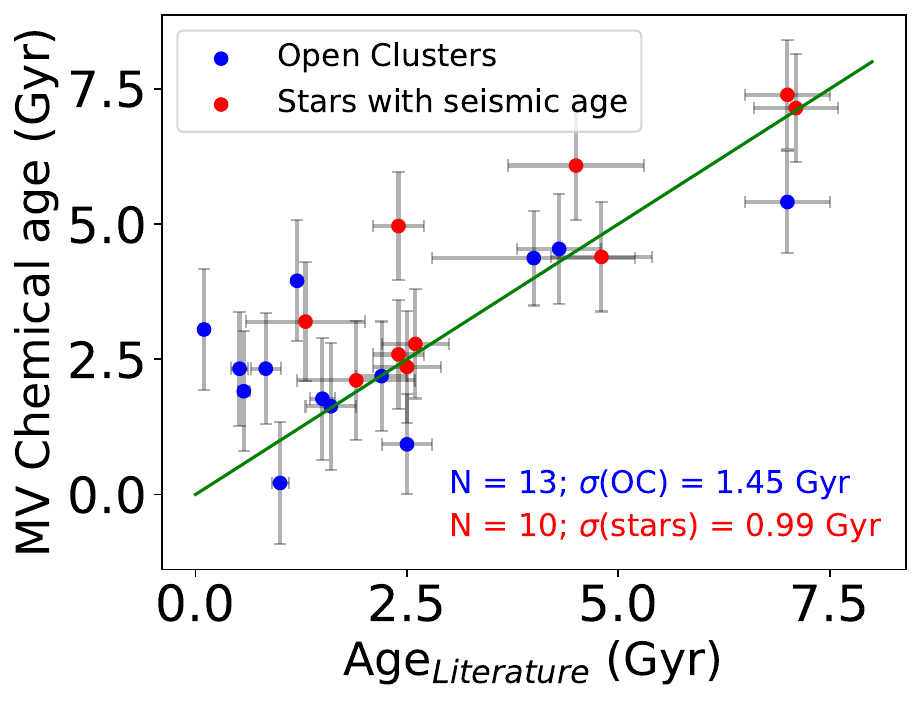} 
\includegraphics[width=\columnwidth]{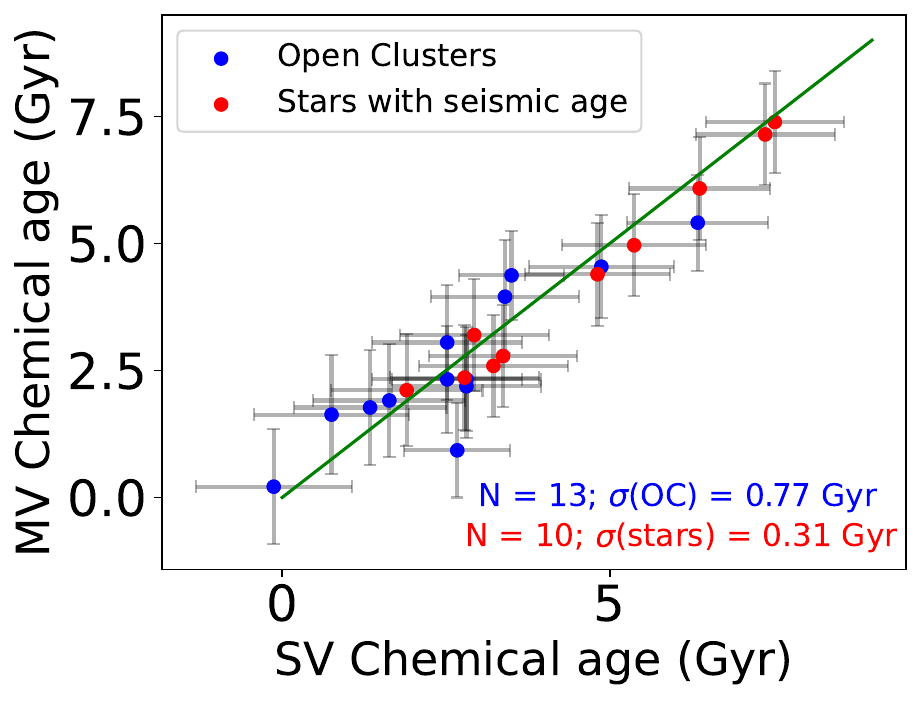} 
\caption{Comparison of the literature ages of the OCs and KLS stars with their SV (upper panel) 
and MV (middle panel) chemical ages. The bottom panel shows the correlation between the two chemical ages. 
The green line represents the one-to-one fit.}\label{Ocs_seismic_age_validation}
\end{figure}

{Figure \ref{Ocs_seismic_age_validation} shows the comparisons of the SV (upper panel) and MV (middle panel) chemical ages of the validation sample calculated 
using our relations with their literature ages. The lower panel of this figure shows the correlation between the 
SV and MV chemical ages of this sample. As can be seen from the figure, the OCs show a larger scatter compared to the KLS stars, but both 
agree with the [Y/Mg] clock within 1$\sigma$ or 2$\sigma$. This may be because 
these cluster members are giants or sub-giants, whereas our relations are derived based on a sample of dwarf stars. 
In addition to this, even though the OCs in the inner Galactic disk (R$_{GC}$ $<$ 6.5 kpc) are excluded from our comparison, 
some of them could have originally belonged to this region and may currently be in the 
solar neighbourhood as a result of radial migration (e.g. \citealt{Viscasillas_2022}). Since our stellar sample lies in the solar neighbourhood, 
this also may have contributed to the scatter seen in Fig. \ref{Ocs_seismic_age_validation}.} 

{Single variate chemical clock (Eq. \ref{equation_dating_relation}): In the case of KLS stars with asteroseismic ages, 
the mean scatter (accuracy) of this chemical clock is 0.91 Gyr, with a 
SD (i.e. precision) of 0.96 Gyr and a mean relative error (MRE) \footnote{RE = $\frac{\lvert Literatute\, age - Chemical\, age \rvert}{Literature\, age}$} of 38\%. 
The average uncertainty in the seismic ages of the sample is $\sim$0.52 Gyr, and the [Y/Mg] has a mean error of 0.013 dex, which translates to $\sim$0.32 Gyr.
By incorporating these two errors in to the mean scatter, the actual accuracy of the SV chemical age is estimated to be $\sim$0.67 Gyr for the KLS stars. 
In the case of OCs, the mean scatter and the SD of the SV chemical clock are 1.10 and 1.24 Gyr, respectively. The average uncertainty in their isochrone ages is 
0.23 Gyr and that from the [Y/Mg] ratio is 0.73 Gyr ($\equiv$ 0.03 dex). Subtracting these errors results in an accuracy of 0.79 Gyr in the SV chemical ages of the OCs.}

{Multivariate chemical clock (Eq. \ref{equation_dating_relation_multivariate}): The mean scatter and the SD are 0.76 Gyr and 0.99 Gyr, respectively, with a 
MRE of 33\% for the KLS stars. In the case of OCs, the mean scatter and the SD for this chemical clock are 1.17 Gyr and 1.45 Gyr, respectively. Incorporating 
the uncertainties in the literature ages and the [Y/Mg] ratios, the accuracy of the MV chemical ages is estimated to be 0.45 Gyr and 0.88 Gyr, respectively, for the
KLS stars and the OCs.}

{In summary, the accuracy we have obtained using our chemical clocks based on the KLS stars with 
asteroseismic ages and the OCs with isochrone ages respectively are $\sim$0.67 and 0.79 Gyr (SV chemical age) and $\sim$0.45 and 0.88 Gyr (MV chemical age). 
These values are similar to the accuracy achieved for the 
solar twin stars in the solar neighbourhood ($\sim$0.8 - 1 Gyr; \citealt{Nissen_2015, Nissen_2016, Tuccimaia_2016, Spina_2018}). Since the KLS stars 
are dwarfs and lie in the close vicinity of the Sun as our stellar sample, the validation using them is likely more accurate compared to the OCs.}

{\footnotesize
\begin{table*}
\centering
\caption{\small {Properties of the OCs and KLS stars used in this study}.} \label{Table_OC_KLS} 
\begin{tabular}{lccc||ccccc} 
\hline                       
ID          &     [Fe/H]         & [Y/Mg]               & Age           & ID          &     [Fe/H]         & [Y/Mg]       & Age       \\  
            &     (dex)          & (dex)                & (Gyr)         &             &     (dex)          & (dex)        & (Gyr)       \\
\hline
            &  OCs$^{\ast}$           &          &               &             &       KLS stars$^{\dagger}$             &          &                     \\
\hline
            &                                   &           &               &              &                                        &           &               \\
Berkeley 31     &       $-$0.270$\pm$0.060      &       $-$0.010$\pm$0.030      &       2.50$\pm$0.30     & KIC 3427720  &       $-$0.024$\pm$0.017      &       $-$0.017$\pm$0.015      &       2.40$\pm$0.30   \\
Berkeley 36     &       $-$0.160$\pm$0.100      &       $-$0.050$\pm$0.060      &       7.00$\pm$0.50     & KIC 6106415  &       $-$0.037$\pm$0.012      &       0.002$\pm$0.011         &       4.80$\pm$0.60   \\
Berkeley 44     &       0.270$\pm$0.060         &       0.140$\pm$0.070         &       1.60$\pm$0.30     & KIC 6225718  &       $-$0.110$\pm$0.014      &       0.051$\pm$0.012         &       2.60$\pm$0.40   \\
M 67            &       $-$0.010$\pm$0.040      &       0.000$\pm$0.010         &       4.30$\pm$0.50     & KIC 7940546  &       $-$0.126$\pm$0.012      &       0.056$\pm$0.010         &       2.40$\pm$0.30   \\
Melotte 71      &       $-$0.090$\pm$0.030      &       0.070$\pm$0.010         &       0.83$\pm$0.18     & KIC 9139151  &       0.096$\pm$0.019         &       0.101$\pm$0.017         &       1.90$\pm$0.70       \\
NGC 2243        &       $-$0.380$\pm$0.040      &       $-$0.040$\pm$0.030      &       4.00$\pm$1.20     & KIC 10162436 &       $-$0.073$\pm$0.021      &       0.071$\pm$0.017         &       2.50$\pm$0.40     \\
NGC 2420        &       $-$0.130$\pm$0.040      &       0.070$\pm$0.030         &       2.20$\pm$0.30     & KIC 10644253 &       0.130$\pm$0.019         &       0.066$\pm$0.017         &       1.30$\pm$0.70     \\
NGC 4815        &       0.110$\pm$0.010         &       0.110$\pm$0.090         &       0.57$\pm$0.07     & KIC 12069424 &       0.093$\pm$0.007         &       $-$0.090$\pm$0.009      &       7.00$\pm$0.50   \\
NGC 6067        &       0.200$\pm$0.080         &       0.080$\pm$0.040         &       0.10$\pm$0.05     & KIC 12069449 &       0.062$\pm$0.007         &       $-$0.085$\pm$0.009      &       7.10$\pm$0.50   \\
NGC 6633        &       $-$0.010$\pm$0.110      &       0.080$\pm$0.020         &       0.52$\pm$0.10     & KIC 12258514 &       0.027$\pm$0.007         &       $-$0.051$\pm$0.016      &       4.50$\pm$0.80   \\
NGC 6802        &       0.100$\pm$0.020         &       0.170$\pm$0.020         &       1.00$\pm$0.10     &                 &                       &                        &                   \\
Pismis 18       &       0.220$\pm$0.040         &       0.050$\pm$0.040         &       1.20$\pm$0.04     &                 &                       &                        &                   \\
Trumpler 20     &       0.150$\pm$0.070         &       0.120$\pm$0.020         &       1.50$\pm$0.15     &                 &                       &                        &                   \\
\hline
\end{tabular} \\ [0.05in]
References: $^{\ast}$ \cite{Casali_2020} and references therein, $^{\dagger}$ \cite{Nissen_2017} and references therein \\
\end{table*} 
}

\subsubsection{Chemical clock in the thin and thick disks}
While the studies using solar twins, analogues, and solar-type stars with [Fe/H] $\geq$ $-$0.3
confirmed the tight correlation between the [Y/Mg] ratio and stellar ages irrespective of the 
population (thin or thick disk; \citealt{Nissen_2016, Tuccimaia_2016, Spina_2016b, Spina_2018, Nissen_2017, Nissen_2020, Jofre_2020}), 
a few recent studies have suggested that the [Y/Mg] chemical clock can only be used for thin disk stars and may
not be valid for lower metallicities or thick disk stars (e.g. \citealt{Feltzing_2017, Delgadomena_2018, Delgadomena_2019, Titarenko_2019, Tautvaivsiene_2021}).
\cite{Feltzing_2017} suggest that the [Y/Mg] - age distribution is almost flat and has no predictive power on age for [Fe/H] $<$ $-$0.5. 
However, \cite{Delgadomena_2018} show that it still has a correlation with age (up to [Fe/H] $\sim$ $-$0.8), but only for thin disk stars. 
The trend becomes flat for thick disk stars, and for thin disk stars with ages $\geq$ 8 Gyr. 
From an AMBRE (Arch{\'e}ologie avec Matisse Bas{\'e}e sur les aRchives de l'ESO) sample of solar-type stars, \cite{Titarenko_2019} identified different slopes for thin disk and thick disk stars with the 
thick disk stars showing a steeper dependence. Recently, \cite{Tautvaivsiene_2021} found that thick disk 
stars (233 stars including dwarfs and giants) show no trend of [Y/Mg] with age (slope$\sim$ $-$0.002), contrarily to the 
slopes identified for the thick disk stars of \cite{Titarenko_2019} and \cite{Bensby_2014}. 
This may be due to the different luminosity classes in their sample.  

\begin{figure}
\centering
\includegraphics[width=\columnwidth]{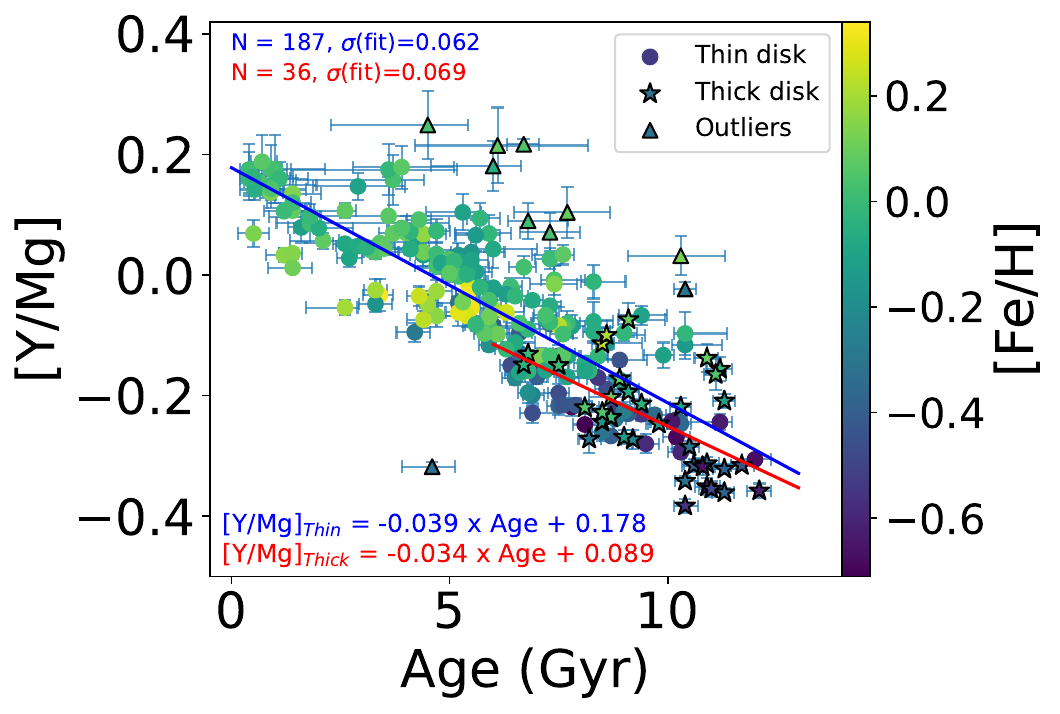} 
\caption{Best linear fits for thin disk stars (solid blue line) and the thick  disk stars (solid red line). The fitting functions are described in the 
text (Eq. \ref{equation_chemical_clock_thin_thick_disk}).}\label{YMg_Thin_Thick_disk}
\end{figure}

We examined the dependence of the [Y/Mg] chemical clock on stellar age for the thin and thick disk stars separately
and found that both components show almost the same correlation, with slopes $-$0.039($\pm$0.001) and $-$0.034($\pm$0.001), respectively. 
The linear fits (Eq. \ref{equation_chemical_clock_thin_thick_disk}) for the two populations are shown in Fig. \ref{YMg_Thin_Thick_disk}. 

\begin{equation}
\label{equation_chemical_clock_thin_thick_disk} 
\begin{split}
    \rm [Y/Mg]_{Thin} = -0.039(\pm0.001) \times \rm Age \,(Gyr) + 0.178(\pm0.038) \\
   \rm [Y/Mg]_{Thick} = -0.034(\pm0.001) \times \rm Age \,(Gyr) + 0.089(\pm0.015). \\   
\end{split}
\end{equation}

\noindent Our results disagree with previous studies, which found that the thick disk stars showed either no trend at all 
or a different trend than the thin disk stars. {In particular, the thick disk stars in \cite{Spina_2018} seem to have a small 
offset relative to the [Y/Mg]--age relation defined by their thin disk sample. 
However, their sample of thick disk stars is relatively small, making it difficult to perform a separate fit for 
the [Y/Mg]--age relation of the thick disk. Although our sample size of thick disk stars has increased relative to the study 
of \cite{Spina_2018}, we have to keep in mind that our number is still relatively small (36 objects without outliers).
We note that the results may be somewhat biased due to the selection biases of thick and thin disk stars in different studies.}

\subsubsection{Comparison with previous works}
The slopes of the [Y/Mg]--age distribution for solar twins and solar-type stars from different studies, along with our estimates, are 
given in Table \ref{Table_slopes_chemical_clock}. These estimates are based on the linear relation between [Y/Mg] and the stellar ages. 
We note that, while the relations obtained for the solar twins by different authors are similar irrespective of the sample size, 
the solar-type stars give different slopes. However, \cite{Nissen_2017}, \cite{Delgadomena_2019}, \cite{Nissen_2020}, and \cite{Tautvaivsiene_2021} show 
similar values that match those of solar twins. Different samples of solar-type stars listed in Table \ref{Table_slopes_chemical_clock} have different 
range of metallicities. Hence, the difference in the slopes plausibly results from the metallicity 
dependence, as discussed in \cite{Feltzing_2017} and \cite{Delgadomena_2018, Delgadomena_2019}. However, when we consider the sample irrespective of the population, 
our relations are compatible with those found for the solar twins by \cite{Nissen_2016} and for solar-type stars by \cite{Nissen_2017,Nissen_2020}. 
While our estimate for the thick disk stars matches the estimate of \cite{Tautvaivsiene_2021} from a sample of 76 thick disk giants,
it disagrees with other samples of thick disk stars. 

The values listed in Table \ref{Table_slopes_chemical_clock} are based on a simple linear regression fit between the chemical clock
and the stellar ages. In light of the results from \cite{Feltzing_2017} and \cite{Delgadomena_2018}, and the non-universality of the 
chemical clock - age relation in the Galactic disk \citep{Casali_2020}, a few recent studies have attempted to perform multiple variable linear 
regression fitting by including, for example, the stellar atmospheric parameters, Galactocentric distances, and stellar masses, in addition to 
the stellar ages \citep{Delgadomena_2019, Casali_2020, Viscasillas_2022}. 

The accuracy and precision of the [Y/Mg] chemical clock when considered as a function of age, [Fe/H], and R$_{GC}$ are 
1.6 Gyr and 2.8 Gyr, respectively \citep{Viscasillas_2022}. \cite{Delgadomena_2019}  were able to determine the ages with a precision of 
1.85 Gyr and MRE$\sim$57\% using their 1D [Y/Mg] chemical clock. When using 2D relations, including T$_{eff}$/[Fe/H]/mass as well,  
they got precisions and MRE (1.57 Gyr, 48\%), (1.73 Gyr, 40\%), and (1.71 Gyr, 37\%), respectively, whereas their 
3D relations (T$_{eff}$ and [Fe/H] along with age) resulted in a precision of 1.47 Gyr with an MRE of 51\%. 
The accuracy achieved by \cite{Titarenko_2019} using their 1D dating relation is 2 Gyr. 
{Meanwhile, the accuracy and precision of the chemical ages calculated using our 
relation (Eq. \ref{equation_dating_relation_multivariate}) are $\sim$0.45 Gyr and 0.99 Gyr, respectively,
with a MRE of 33\%.}
Since our sample lies in the solar neighbourhood, we have only discussed the results valid in this region.

{\footnotesize
\begin{table*}
\caption{Slopes of the [Y/Mg]--age distribution of stars in the solar neighbourhood.} \label{Table_slopes_chemical_clock}
\resizebox{\textwidth}{!}{
\begin{tabular}{lccc||cccccc}
\hline                       
Slope                       & N$^{\dagger}$         & Pop$^{\ast}$                          & Ref    &  Slope                       & N$^{\dagger}$         & Pop$^{\ast}$              & [Fe/H]         & Ref     \\ 
(dex Gyr$^{-1}$)            &                       &                                                    &    & (dex Gyr$^{-1}$)            &                       &                 &               &       \\  
\hline
                            &           &                                         &                       &                              &           &                  &                       &                    \\
solar twins                 &           &                                         &                       & solar-type                  &           &                  &                       &                    \\
                            &           &                                         &                       &                              &           &                  &                       &                    \\
$-$0.0404$\pm$0.0019        & 18        & Thin disk                               &  \cite{Nissen_2015}   & $-$0.0362$\pm$0.0004         & 57       &                  &  $-$0.71 - $-$0.17      & This work        \\
$-$0.0371$\pm$0.0013        & 21        &                                         & \cite{Nissen_2016}    & $-$0.0340$\pm$0.0006         & 166      &                  & $-$0.17 - +0.34        & This work          \\
$-$0.0410$\pm$0.0010        & 65        &                                         & \cite{Tuccimaia_2016} & $-$0.0390$\pm$0.0010         & 187       & Thin disk        & $-$0.71 - +0.34        & This work        \\
$-$0.0410$\pm$0.0017        & 45        &                                         & \cite{Spina_2016b}    & $-$0.0340$\pm$0.0010         & 36       & Thick disk       & $-$0.71 - +0.05        & This work      \\
$-$0.0460$\pm$0.0020        & 76        &                                         & \cite{Spina_2018}    & $-$0.0347$\pm$0.0012        & 31        &                  & $-$0.15 - +0.15       & \cite{Nissen_2017}  \\
$-$0.0420$\pm$0.0030        & 66        & Thin disk                               & \cite{Spina_2018}     & $-$0.0199                   &           & Thin disk        & $-$0.80 - $-$0.20       & \cite{Delgadomena_2018} \\
$-$0.0410$\pm$0.0030        &           &                                         & \cite{Delgadomena_2019}   & $-$0.0420$\pm$0.0010        & 354       &         & $-$1.15 - +0.55        & \cite{Delgadomena_2019}       \\
$-$0.0420$\pm$0.0030        & 80        &                                         & \cite{Jofre_2020}   & $-$0.0290$\pm$0.0020        & 22        & Thin disk        & $-$0.80 - +0.40       & \cite{Titarenko_2019}     \\
$-$0.0400$\pm$0.0020        &           &                                         & \cite{Casali_2020}       & $-$0.0820$\pm$0.0020         & 11        & Thick disk       & $-$0.80 - +0.20       & \cite{Titarenko_2019}     \\
                            &           &                                         &                         & $-$0.0383$\pm$0.0010        & 68        &                  & $-$0.30 - +0.30      & \cite{Nissen_2020}\\
                            &           &                                         &                         & $-$0.0270$\pm$0.0030        & 368       & Thin disk        & $-$0.60 - +0.40    & \cite{Tautvaivsiene_2021}$^{\ddagger}$ \\
                            &           &                                         &                         & $-$0.0410$\pm$0.0130$^{\star}$          & 76        & Thick disk       & $-$0.46$\pm$0.12      &           \cite{Tautvaivsiene_2021} \\
                            &           &                                         &                         & $-$0.0140$\pm$0.0040        & 66        & Thick disk       &  $-$1.00 - +0.40                  & \cite{Tautvaivsiene_2021} (for \cite{Bensby_2014} sample) \\
                            &           &                                         &                         & $-$0.0020$\pm$0.0030        & 233       & Thick disk       &  $-$1.00 - +0.40                  &    
 \cite{Tautvaivsiene_2021} and references therein$^{\ddagger}$                \\                         
\hline
\end{tabular} } \\
$^{\dagger}$ No. of stars used to derive the relation \\
$^{\ast}$ Fitting includes both the populations if not mentioned \\
$^{\ddagger}$ Sample includes dwarf and giants \\
$^{\star}$ Only giants  \\
\end{table*}
}

\section{Conclusions}  \label{section_conclusion}
Based on high-resolution and high-quality HARPS spectra, we estimated the precise stellar atmospheric parameters and
chemical abundances of Mg and Y through a line-by-line differential analysis for a sample of 48 metal-poor solar-type stars in the 
solar neighbourhood. Through the isochrone fitting method, the stellar masses and ages of the sample were also estimated. 
We performed a joint analysis of these 48 objects, together with 185 solar twins and analogues from our previous studies. 
The combined sample consists of 233 solar-type stars in the solar neighbourhood with metallicities in the range $-$0.71 $\leq$ [Fe/H] $<$ +0.34. 
The main results of our homogeneous analysis are summarised as follows:  

\begin{itemize}
    \item The sample shows a clear bimodality in the [Mg/Fe]--[Fe/H] plane, and the thin and thick disk populations are clearly distinguishable. 
    Thick disk stars show higher [$\alpha$/Fe] values compared to thin disk stars. While the thick disk stars are older than 8 Gyr, 
    the thin disk shows an age spread. Nonetheless, there is no age discontinuity between the two populations.
    \item {Our chemically defined thick disk and thin disk stars overlap in the kinematic space, a behaviour already noted in
    several previous studies.}
    \item The thick disk stars show an AMR; however, the age and metallicity exhibit an essentially flat distribution and are uncorrelated
       for the thin disk stars. On the other hand, the two populations blend together in the age--[Fe/H] distribution.
       \item There exists a correlation between [Mg/Fe] and stellar ages for both populations. However, in the case of [Y/Fe], only the thin disk stars show a correlation.
       We note a clear dichotomy between these two populations in the [Mg/Fe]--age distribution, whereas there is no separation between them in the [Y/Fe]--age space. 
       \item The [Y/Mg] ratio shows a strong anti-correlation with stellar age. Our analysis confirms the metallicity dependence of the [Y/Mg] chemical clock. 
       We note the scatter in our [Y/Mg]--age distribution is lower than that of all previous studies of solar-type stars. There is no discontinuity between the thin 
       and thick disks in the [Y/Mg]--age correlation. Our estimates of the slope are compatible with those found for solar twins 
       and other samples of solar-type stars in the higher metallicity bin ([Fe/H] $>$ $-$0.30). 
       \item For the first time in the literature, we report similar slopes of the [Y/Mg]--age correlation for thin ($-$0.039) and thick disk ($-$0.034) stars. 
       \item {Using the dating relations (Eqs. \ref{equation_dating_relation} and \ref{equation_dating_relation_multivariate}), 
           we achieve high accuracies (0.67 and 0.45 Gyr, respectively) and precisions (0.96 and 0.99 Gyr, respectively). 
           These accuracies are compatible with the best accuracy achieved for solar twins to date ($\sim$0.8 - 1.0 Gyr) in the solar neighbourhood.
           The MREs in the chemical ages using these relations are 38\% and 33\%, respectively. The chemical clock based on the MV relation (Eq. \ref{equation_dating_relation_multivariate}) 
           performs somewhat better than the SV relations in two metallicity regimes (Eq. \ref{equation_dating_relation}). }
\end{itemize}

The relations connecting the abundance ratios and stellar ages are related to the chemo-dynamical evolution of the 
Galaxy and can be used to study different components of it \citep{Haywood_2013, Haywood_2016}. 
From our analysis, we find that the [Y/Mg] chemical clock 
can be used as an age proxy for solar-type stars. The empirical relations showcased in this work should be
used only for stars in the solar neighbourhood and in the same parameter range as our sample. 
The high-precision outcomes of our work are expected to provide contributions to 
the understanding of different populations of the Galaxy and its chemical evolution.

\begin{acknowledgements}
JS and JM acknowledge the support from FAPESP (2022/10325-3 and 2018/04055-8). 
This work made use of the SIMBAD astronomical database, operated
at CDS, Strasbourg, France, and the NASA ADS, USA.
This work has made use of data from the European Space Agency (ESA) mission
{\it Gaia} (\url{https://www.cosmos.esa.int/gaia}), processed by the {\it Gaia}
Data Processing and Analysis Consortium (DPAC,
\url{https://www.cosmos.esa.int/web/gaia/dpac/consortium}). Funding for the DPAC
has been provided by national institutions, in particular the institutions
participating in the {\it Gaia} Multilateral Agreement.

Based on data collected at the European Southern Observatory under ESO programs
60.A-9036(A), 60.A-9700(G), 60.A-9709(G), 072.C-0488(E), 073.C-0733(A), 077.C-0295(A), 077.C-0295(B), 
078.C-0209(A), 080.C-0712(A), 082.C-0212(A), 082.C-0212(B), 084.C-0229(A), 085.C-0019(A), 
085.C-0063(A), 086.C-0284(A), 087.C-0831(A), 088.C-0323(A), 089.C-0732(A), 091.C-0034(A), 091.C-0936(A), 
092.C-0579(A), 092.C-0721(A), 093.C-0409(A), 094.C-0797(A), 096.C-0053(A), 096.C-0210(A), 098.C-0366(A), 
106.21TJ.001, 183.C-0972(A), 184.C-0815(A), 184.C-0815(C), 184.C-0815(F), 190.C-0027(A), 
192.C-0852(A), 196.C-0042, 196.C-0042(D), 196.C-0042(E), 0100.C-0097(A), 1102.C-0923(A), 
and 1102.C-0923(C).
\end{acknowledgements}

\bibliography{Bibliography}

\begin{thebibliography}{103}
\expandafter\ifx\csname natexlab\endcsname\relax\def\natexlab#1{#1}\fi

\bibitem[{{Adibekyan} {et~al.}(2013){Adibekyan}, {Figueira}, {Santos},
  {Hakobyan}, {Sousa}, {Pace}, {Delgado Mena}, {Robin}, {Israelian}, \&
  {Gonz{\'a}lez Hern{\'a}ndez}}]{Adibekyen_2013}
{Adibekyan}, V.~Z., {Figueira}, P., {Santos}, N.~C., {et~al.} 2013, \aap, 554,
  A44

\bibitem[{{Adibekyan} {et~al.}(2011){Adibekyan}, {Santos}, {Sousa}, \&
  {Israelian}}]{Adibekyan_2011}
{Adibekyan}, V.~Z., {Santos}, N.~C., {Sousa}, S.~G., \& {Israelian}, G. 2011,
  \aap, 535, L11

\bibitem[{{Adibekyan} {et~al.}(2012){Adibekyan}, {Sousa}, {Santos}, {Delgado
  Mena}, {Gonz{\'a}lez Hern{\'a}ndez}, {Israelian}, {Mayor}, \&
  {Khachatryan}}]{Adibekyan_2012}
{Adibekyan}, V.~Z., {Sousa}, S.~G., {Santos}, N.~C., {et~al.} 2012, \aap, 545,
  A32

\bibitem[{{Alfredo Collazos}(2023)}]{Alfred_2023}
{Alfredo Collazos}, J. 2023, arXiv e-prints, arXiv:2308.08492

\bibitem[{{Anders} {et~al.}(2014){Anders}, {Chiappini}, {Santiago},
  {Rocha-Pinto}, {Girardi}, {da Costa}, {Maia}, {Steinmetz}, {Minchev},
  {Schultheis}, {Boeche}, {Miglio}, {Montalb{\'a}n}, {Schneider}, {Beers},
  {Cunha}, {Allende Prieto}, {Balbinot}, {Bizyaev}, {Brauer}, {Brinkmann},
  {Frinchaboy}, {Garc{\'\i}a P{\'e}rez}, {Hayden}, {Hearty}, {Holtzman},
  {Johnson}, {Kinemuchi}, {Majewski}, {Malanushenko}, {Malanushenko},
  {Nidever}, {O'Connell}, {Pan}, {Robin}, {Schiavon}, {Shetrone}, {Skrutskie},
  {Smith}, {Stassun}, \& {Zasowski}}]{Anders_2014}
{Anders}, F., {Chiappini}, C., {Santiago}, B.~X., {et~al.} 2014, \aap, 564,
  A115

\bibitem[{{Babusiaux} {et~al.}(2022){Babusiaux}, {Fabricius}, {Khanna},
  {Muraveva}, {Reyl{\'e}}, {Spoto}, {Vallenari}, {Luri}, {Arenou}, {Alvarez},
  {Anders}, {Antoja}, {Balbinot}, {Barache}, {Bauchet}, {Bossini}, {Busonero},
  {Cantat-Gaudin}, {Carrasco}, {Dafonte}, {Diakite}, {Figueras},
  {Garcia-Gutierrez}, {Garofalo}, {Helmi}, {Jimenez-Arranz}, {Jordi},
  {Kervella}, {Kostrzewa-Rutkowska}, {Leclerc}, {Licata}, {Manteiga}, {Masip},
  {Monguio}, {Ramos}, {Robichon}, {Robin}, {Romero-Gomez}, {Saez}, {Santovena},
  {Spina}, {Torralba Elipe}, \& {Weiler}}]{Babusiaux_2022}
{Babusiaux}, C., {Fabricius}, C., {Khanna}, S., {et~al.} 2022, arXiv e-prints,
  arXiv:2206.05989

\bibitem[{{Bedell} {et~al.}(2014){Bedell}, {Mel{\'e}ndez}, {Bean},
  {Ram{\'\i}rez}, {Leite}, \& {Asplund}}]{Bedell_2014}
{Bedell}, M., {Mel{\'e}ndez}, J., {Bean}, J.~L., {et~al.} 2014, \apj, 795, 23

\bibitem[{{Bensby} {et~al.}(2003){Bensby}, {Feltzing}, \&
  {Lundstr{\"o}m}}]{Bensby_2003}
{Bensby}, T., {Feltzing}, S., \& {Lundstr{\"o}m}, I. 2003, \aap, 410, 527

\bibitem[{{Bensby} {et~al.}(2005){Bensby}, {Feltzing}, {Lundstr{\"o}m}, \&
  {Ilyin}}]{Bensby_2005}
{Bensby}, T., {Feltzing}, S., {Lundstr{\"o}m}, I., \& {Ilyin}, I. 2005, \aap,
  433, 185

\bibitem[{{Bensby} {et~al.}(2014){Bensby}, {Feltzing}, \& {Oey}}]{Bensby_2014}
{Bensby}, T., {Feltzing}, S., \& {Oey}, M.~S. 2014, \aap, 562, A71

\bibitem[{{Bergemann} {et~al.}(2014){Bergemann}, {Ruchti}, {Serenelli},
  {Feltzing}, {Alves-Brito}, {Asplund}, {Bensby}, {Gruyters}, {Heiter},
  {Hourihane}, {Korn}, {Lind}, {Marino}, {Jofre}, {Nordlander}, {Ryde},
  {Worley}, {Gilmore}, {Randich}, {Ferguson}, {Jeffries}, {Micela},
  {Negueruela}, {Prusti}, {Rix}, {Vallenari}, {Alfaro}, {Allende Prieto},
  {Bragaglia}, {Koposov}, {Lanzafame}, {Pancino}, {Recio-Blanco}, {Smiljanic},
  {Walton}, {Costado}, {Franciosini}, {Hill}, {Lardo}, {de Laverny}, {Magrini},
  {Maiorca}, {Masseron}, {Morbidelli}, {Sacco}, {Kordopatis}, \&
  {Tautvai{\v{s}}ien{\.{e}}}}]{Bergemann_2014}
{Bergemann}, M., {Ruchti}, G.~R., {Serenelli}, A., {et~al.} 2014, \aap, 565,
  A89

\bibitem[{{Bland-Hawthorn} \& {Gerhard}(2016)}]{Hawthorn_2016}
{Bland-Hawthorn}, J. \& {Gerhard}, O. 2016, \araa, 54, 529

\bibitem[{{Bovy}(2015)}]{Bovy_2015}
{Bovy}, J. 2015, \apjs, 216, 29

\bibitem[{{Bovy} {et~al.}(2012{\natexlab{a}}){Bovy}, {Rix}, \&
  {Hogg}}]{Bovy_2012a}
{Bovy}, J., {Rix}, H.-W., \& {Hogg}, D.~W. 2012{\natexlab{a}}, \apj, 751, 131

\bibitem[{{Bovy} {et~al.}(2012{\natexlab{b}}){Bovy}, {Rix}, {Liu}, {Hogg},
  {Beers}, \& {Lee}}]{Bovy_2012b}
{Bovy}, J., {Rix}, H.-W., {Liu}, C., {et~al.} 2012{\natexlab{b}}, \apj, 753,
  148

\bibitem[{{Cantelli} \& {Teixeira}(2024)}]{Cantelli_2024}
{Cantelli}, E. \& {Teixeira}, R. 2024, \mnras, 530, 2648

\bibitem[{{Casagrande} {et~al.}(2011){Casagrande}, {Sch{\"o}nrich}, {Asplund},
  {Cassisi}, {Ram{\'\i}rez}, {Mel{\'e}ndez}, {Bensby}, \&
  {Feltzing}}]{Casagrande_2011}
{Casagrande}, L., {Sch{\"o}nrich}, R., {Asplund}, M., {et~al.} 2011, \aap, 530,
  A138

\bibitem[{{Casali} {et~al.}(2019){Casali}, {Magrini}, {Tognelli}, {Jackson},
  {Jeffries}, {Lagarde}, {Tautvai{\v{s}}ien{\.{e}}}, {Masseron},
  {Degl'Innocenti}, {Prada Moroni}, {Kordopatis}, {Pancino}, {Randich},
  {Feltzing}, {Sahlholdt}, {Spina}, {Friel}, {Roccatagliata}, {Sanna},
  {Bragaglia}, {Drazdauskas}, {Mikolaitis}, {Minkevi{\v{c}}i{\={u}}t{\.{e}}},
  {Stonkut{\.{e}}}, {Chorniy}, {Bagdonas}, {Jimenez-Esteban}, {Martell}, {Van
  der Swaelmen}, {Gilmore}, {Vallenari}, {Bensby}, {Koposov}, {Korn}, {Worley},
  {Smiljanic}, {Bergemann}, {Carraro}, {Damiani}, {Prisinzano}, {Bonito},
  {Franciosini}, {Gonneau}, {Hourihane}, {Jofre}, {Lewis}, {Morbidelli},
  {Sacco}, {Sousa}, {Zaggia}, {Lanzafame}, {Heiter}, {Frasca}, \&
  {Bayo}}]{Casali_2019}
{Casali}, G., {Magrini}, L., {Tognelli}, E., {et~al.} 2019, \aap, 629, A62

\bibitem[{{Casali} {et~al.}(2020){Casali}, {Spina}, {Magrini}, {Karakas},
  {Kobayashi}, {Casey}, {Feltzing}, {Van der Swaelmen}, {Tsantaki},
  {Jofr{\'e}}, {Bragaglia}, {Feuillet}, {Bensby}, {Biazzo}, {Gonneau},
  {Tautvai{\v{s}}ien{\.{e}}}, {Baratella}, {Roccatagliata}, {Pancino}, {Sousa},
  {Adibekyan}, {Martell}, {Bayo}, {Jackson}, {Jeffries}, {Gilmore}, {Randich},
  {Alfaro}, {Koposov}, {Korn}, {Recio-Blanco}, {Smiljanic}, {Franciosini},
  {Hourihane}, {Monaco}, {Morbidelli}, {Sacco}, {Worley}, \&
  {Zaggia}}]{Casali_2020}
{Casali}, G., {Spina}, L., {Magrini}, L., {et~al.} 2020, \aap, 639, A127

\bibitem[{{Casamiquela} {et~al.}(2021){Casamiquela}, {Soubiran}, {Jofr{\'e}},
  {Chiappini}, {Lagarde}, {Tarricq}, {Carrera}, {Jordi},
  {Balaguer-N{\'u}{\~n}ez}, {Carbajo-Hijarrubia}, \&
  {Blanco-Cuaresma}}]{Casamiquela_2021}
{Casamiquela}, L., {Soubiran}, C., {Jofr{\'e}}, P., {et~al.} 2021, \aap, 652,
  A25

\bibitem[{{Castelli} \& {Kurucz}(2003)}]{Castelli_2003}
{Castelli}, F. \& {Kurucz}, R.~L. 2003, in Proceedings of the IAU Symp. No 210,
  Vol. 210, Modelling of Stellar Atmospheres, ed. N.~{Piskunov}, W.~W. {Weiss},
  \& D.~F. {Gray}, A20

\bibitem[{{Cayrel de Strobel}(1996)}]{Cayrel_1996}
{Cayrel de Strobel}, G. 1996, \aapr, 7, 243

\bibitem[{{Chaplin} \& {Miglio}(2013)}]{Chaplin_2013}
{Chaplin}, W.~J. \& {Miglio}, A. 2013, \araa, 51, 353

\bibitem[{{Christensen-Dalsgaard}(2016)}]{Christensen_2016}
{Christensen-Dalsgaard}, J. 2016, arXiv e-prints, arXiv:1602.06838

\bibitem[{{Cox}(2000)}]{Cox_2000}
{Cox}, A.~N. 2000, {Allen's astrophysical quantities} (Springer)

\bibitem[{{da Silva} {et~al.}(2012){da Silva}, {Porto de Mello}, {Milone}, {da
  Silva}, {Ribeiro}, \& {Rocha-Pinto}}]{dasilva_2012}
{da Silva}, R., {Porto de Mello}, G.~F., {Milone}, A.~C., {et~al.} 2012, \aap,
  542, A84

\bibitem[{{Delgado Mena} {et~al.}(2019){Delgado Mena}, {Moya}, {Adibekyan},
  {Tsantaki}, {Gonz{\'a}lez Hern{\'a}ndez}, {Israelian}, {Davies}, {Chaplin},
  {Sousa}, {Ferreira}, \& {Santos}}]{Delgadomena_2019}
{Delgado Mena}, E., {Moya}, A., {Adibekyan}, V., {et~al.} 2019, \aap, 624, A78

\bibitem[{{Delgado Mena} {et~al.}(2018){Delgado Mena}, {Tsantaki}, {Zh.
  Adibekyan}, {Sousa}, {Santos}, {Gonz{\'a}lez Hern{\'a}ndez}, \&
  {Israelian}}]{Delgadomena_2018}
{Delgado Mena}, E., {Tsantaki}, M., {Zh. Adibekyan}, V., {et~al.} 2018, in
  Astrometry and Astrophysics in the Gaia Sky, ed. A.~{Recio-Blanco}, P.~{de
  Laverny}, A.~G.~A. {Brown}, \& T.~{Prusti}, Vol. 330, 156--159

\bibitem[{{Edvardsson} {et~al.}(1993){Edvardsson}, {Andersen}, {Gustafsson},
  {Lambert}, {Nissen}, \& {Tomkin}}]{Edvardsson_1993}
{Edvardsson}, B., {Andersen}, J., {Gustafsson}, B., {et~al.} 1993, \aap, 275,
  101

\bibitem[{{Epstein} {et~al.}(2010){Epstein}, {Johnson}, {Dong}, {Udalski},
  {Gould}, \& {Becker}}]{Epstein_2010}
{Epstein}, C.~R., {Johnson}, J.~A., {Dong}, S., {et~al.} 2010, \apj, 709, 447

\bibitem[{{Feltzing} {et~al.}(2001){Feltzing}, {Holmberg}, \&
  {Hurley}}]{Feltzing_2001}
{Feltzing}, S., {Holmberg}, J., \& {Hurley}, J.~R. 2001, \aap, 377, 911

\bibitem[{{Feltzing} {et~al.}(2017){Feltzing}, {Howes}, {McMillan}, \&
  {Stonkut{\.{e}}}}]{Feltzing_2017}
{Feltzing}, S., {Howes}, L.~M., {McMillan}, P.~J., \& {Stonkut{\.{e}}}, E.
  2017, \mnras, 465, L109

\bibitem[{{Fuhrmann}(1998)}]{Fuhrmann_1998}
{Fuhrmann}, K. 1998, \aap, 338, 161

\bibitem[{{Fuhrmann} {et~al.}(2012){Fuhrmann}, {Chini}, {Haas}, {Hackstein},
  {Ramolla}, \& {Bernkopf}}]{Fuhrmann_2012}
{Fuhrmann}, K., {Chini}, R., {Haas}, M., {et~al.} 2012, \apj, 761, 159

\bibitem[{{Gaia Collaboration} {et~al.}(2016){Gaia Collaboration}, {Prusti},
  {de Bruijne}, {Brown}, {Vallenari}, {Babusiaux}, {Bailer-Jones}, {Bastian},
  {Biermann}, {Evans}, {Eyer}, {Jansen}, {Jordi}, {Klioner}, {Lammers},
  {Lindegren}, {Luri}, {Mignard}, {Milligan}, {Panem}, {Poinsignon},
  {Pourbaix}, {Randich}, {Sarri}, {Sartoretti}, {Siddiqui}, {Soubiran},
  {Valette}, {van Leeuwen}, {Walton}, {Aerts}, {Arenou}, {Cropper}, {Drimmel},
  {H{\o}g}, {Katz}, {Lattanzi}, {O'Mullane}, {Grebel}, {Holland}, {Huc},
  {Passot}, {Bramante}, {Cacciari}, {Casta{\~n}eda}, {Chaoul}, {Cheek}, {De
  Angeli}, {Fabricius}, {Guerra}, {Hern{\'a}ndez}, {Jean-Antoine-Piccolo},
  {Masana}, {Messineo}, {Mowlavi}, {Nienartowicz}, {Ord{\'o}{\~n}ez-Blanco},
  {Panuzzo}, {Portell}, {Richards}, {Riello}, {Seabroke}, {Tanga},
  {Th{\'e}venin}, {Torra}, {Els}, {Gracia-Abril}, {Comoretto},
  {Garcia-Reinaldos}, {Lock}, {Mercier}, {Altmann}, {Andrae}, {Astraatmadja},
  {Bellas-Velidis}, {Benson}, {Berthier}, {Blomme}, {Busso}, {Carry},
  {Cellino}, {Clementini}, {Cowell}, {Creevey}, {Cuypers}, {Davidson}, {De
  Ridder}, {de Torres}, {Delchambre}, {Dell'Oro}, {Ducourant}, {Fr{\'e}mat},
  {Garc{\'\i}a-Torres}, {Gosset}, {Halbwachs}, {Hambly}, {Harrison}, {Hauser},
  {Hestroffer}, {Hodgkin}, {Huckle}, {Hutton}, {Jasniewicz}, {Jordan},
  {Kontizas}, {Korn}, {Lanzafame}, {Manteiga}, {Moitinho}, {Muinonen},
  {Osinde}, {Pancino}, {Pauwels}, {Petit}, {Recio-Blanco}, {Robin}, {Sarro},
  {Siopis}, {Smith}, {Smith}, {Sozzetti}, {Thuillot}, {van Reeven}, {Viala},
  {Abbas}, {Abreu Aramburu}, {Accart}, {Aguado}, {Allan}, {Allasia},
  {Altavilla}, {{\'A}lvarez}, {Alves}, {Anderson}, {Andrei}, {Anglada Varela},
  {Antiche}, {Antoja}, {Ant{\'o}n}, {Arcay}, {Atzei}, {Ayache}, {Bach},
  {Baker}, {Balaguer-N{\'u}{\~n}ez}, {Barache}, {Barata}, {Barbier}, {Barblan},
  {Baroni}, {Barrado y Navascu{\'e}s}, {Barros}, {Barstow}, {Becciani},
  {Bellazzini}, {Bellei}, {Bello Garc{\'\i}a}, {Belokurov}, {Bendjoya},
  {Berihuete}, {Bianchi}, {Bienaym{\'e}}, {Billebaud}, {Blagorodnova},
  {Blanco-Cuaresma}, {Boch}, {Bombrun}, {Borrachero}, {Bouquillon}, {Bourda},
  {Bouy}, {Bragaglia}, {Breddels}, {Brouillet}, {Br{\"u}semeister},
  {Bucciarelli}, {Budnik}, {Burgess}, {Burgon}, {Burlacu}, {Busonero}, {Buzzi},
  {Caffau}, {Cambras}, {Campbell}, {Cancelliere}, {Cantat-Gaudin}, {Carlucci},
  {Carrasco}, {Castellani}, {Charlot}, {Charnas}, {Charvet}, {Chassat},
  {Chiavassa}, {Clotet}, {Cocozza}, {Collins}, {Collins}, {Costigan}, {Crifo},
  {Cross}, {Crosta}, {Crowley}, {Dafonte}, {Damerdji}, {Dapergolas}, {David},
  {David}, {De Cat}, {de Felice}, {de Laverny}, {De Luise}, {De March}, {de
  Martino}, {de Souza}, {Debosscher}, {del Pozo}, {Delbo}, {Delgado},
  {Delgado}, {di Marco}, {Di Matteo}, {Diakite}, {Distefano}, {Dolding}, {Dos
  Anjos}, {Drazinos}, {Dur{\'a}n}, {Dzigan}, {Ecale}, {Edvardsson}, {Enke},
  {Erdmann}, {Escolar}, {Espina}, {Evans}, {Eynard Bontemps}, {Fabre},
  {Fabrizio}, {Faigler}, {Falc{\~a}o}, {Farr{\`a}s Casas}, {Faye}, {Federici},
  {Fedorets}, {Fern{\'a}ndez-Hern{\'a}ndez}, {Fernique}, {Fienga}, {Figueras},
  {Filippi}, {Findeisen}, {Fonti}, {Fouesneau}, {Fraile}, {Fraser}, {Fuchs},
  {Furnell}, {Gai}, {Galleti}, {Galluccio}, {Garabato}, {Garc{\'\i}a-Sedano},
  {Gar{\'e}}, {Garofalo}, {Garralda}, {Gavras}, {Gerssen}, {Geyer}, {Gilmore},
  {Girona}, {Giuffrida}, {Gomes}, {Gonz{\'a}lez-Marcos},
  {Gonz{\'a}lez-N{\'u}{\~n}ez}, {Gonz{\'a}lez-Vidal}, {Granvik}, {Guerrier},
  {Guillout}, {Guiraud}, {G{\'u}rpide}, {Guti{\'e}rrez-S{\'a}nchez}, {Guy},
  {Haigron}, {Hatzidimitriou}, {Haywood}, {Heiter}, {Helmi}, {Hobbs},
  {Hofmann}, {Holl}, {Holland}, {Hunt}, {Hypki}, {Icardi}, {Irwin}, {Jevardat
  de Fombelle}, {Jofr{\'e}}, {Jonker}, {Jorissen}, {Julbe}, {Karampelas},
  {Kochoska}, {Kohley}, {Kolenberg}, {Kontizas}, {Koposov}, {Kordopatis},
  {Koubsky}, {Kowalczyk}, {Krone-Martins}, {Kudryashova}, {Kull}, {Bachchan},
  {Lacoste-Seris}, {Lanza}, {Lavigne}, {Le Poncin-Lafitte}, {Lebreton},
  {Lebzelter}, {Leccia}, {Leclerc}, {Lecoeur-Taibi}, {Lemaitre}, {Lenhardt},
  {Leroux}, {Liao}, {Licata}, {Lindstr{\o}m}, {Lister}, {Livanou}, {Lobel},
  {L{\"o}ffler}, {L{\'o}pez}, {Lopez-Lozano}, {Lorenz}, {Loureiro},
  {MacDonald}, {Magalh{\~a}es Fernandes}, {Managau}, {Mann}, {Mantelet},
  {Marchal}, {Marchant}, {Marconi}, {Marie}, {Marinoni}, {Marrese},
  {Marschalk{\'o}}, {Marshall}, {Mart{\'\i}n-Fleitas}, {Martino}, {Mary},
  {Matijevi{\v{c}}}, {Mazeh}, {McMillan}, {Messina}, {Mestre}, {Michalik},
  {Millar}, {Miranda}, {Molina}, {Molinaro}, {Molinaro}, {Moln{\'a}r},
  {Moniez}, {Montegriffo}, {Monteiro}, {Mor}, {Mora}, {Morbidelli}, {Morel},
  {Morgenthaler}, {Morley}, {Morris}, {Mulone}, {Muraveva}, {Musella},
  {Narbonne}, {Nelemans}, {Nicastro}, {Noval}, {Ord{\'e}novic},
  {Ordieres-Mer{\'e}}, {Osborne}, {Pagani}, {Pagano}, {Pailler}, {Palacin},
  {Palaversa}, {Parsons}, {Paulsen}, {Pecoraro}, {Pedrosa}, {Pentik{\"a}inen},
  {Pereira}, {Pichon}, {Piersimoni}, {Pineau}, {Plachy}, {Plum}, {Poujoulet},
  {Pr{\v{s}}a}, {Pulone}, {Ragaini}, {Rago}, {Rambaux}, {Ramos-Lerate},
  {Ranalli}, {Rauw}, {Read}, {Regibo}, {Renk}, {Reyl{\'e}}, {Ribeiro},
  {Rimoldini}, {Ripepi}, {Riva}, {Rixon}, {Roelens}, {Romero-G{\'o}mez},
  {Rowell}, {Royer}, {Rudolph}, {Ruiz-Dern}, {Sadowski}, {Sagrist{\`a}
  Sell{\'e}s}, {Sahlmann}, {Salgado}, {Salguero}, {Sarasso}, {Savietto},
  {Schnorhk}, {Schultheis}, {Sciacca}, {Segol}, {Segovia}, {Segransan},
  {Serpell}, {Shih}, {Smareglia}, {Smart}, {Smith}, {Solano}, {Solitro},
  {Sordo}, {Soria Nieto}, {Souchay}, {Spagna}, {Spoto}, {Stampa}, {Steele},
  {Steidelm{\"u}ller}, {Stephenson}, {Stoev}, {Suess}, {S{\"u}veges}, {Surdej},
  {Szabados}, {Szegedi-Elek}, {Tapiador}, {Taris}, {Tauran}, {Taylor},
  {Teixeira}, {Terrett}, {Tingley}, {Trager}, {Turon}, {Ulla}, {Utrilla},
  {Valentini}, {van Elteren}, {Van Hemelryck}, {van Leeuwen}, {Varadi},
  {Vecchiato}, {Veljanoski}, {Via}, {Vicente}, {Vogt}, {Voss}, {Votruba},
  {Voutsinas}, {Walmsley}, {Weiler}, {Weingrill}, {Werner}, {Wevers},
  {Whitehead}, {Wyrzykowski}, {Yoldas}, {{\v{Z}}erjal}, {Zucker}, {Zurbach},
  {Zwitter}, {Alecu}, {Allen}, {Allende Prieto}, {Amorim},
  {Anglada-Escud{\'e}}, {Arsenijevic}, {Azaz}, {Balm}, {Beck}, {Bernstein},
  {Bigot}, {Bijaoui}, {Blasco}, {Bonfigli}, {Bono}, {Boudreault}, {Bressan},
  {Brown}, {Brunet}, {Bunclark}, {Buonanno}, {Butkevich}, {Carret}, {Carrion},
  {Chemin}, {Ch{\'e}reau}, {Corcione}, {Darmigny}, {de Boer}, {de Teodoro}, {de
  Zeeuw}, {Delle Luche}, {Domingues}, {Dubath}, {Fodor}, {Fr{\'e}zouls},
  {Fries}, {Fustes}, {Fyfe}, {Gallardo}, {Gallegos}, {Gardiol}, {Gebran},
  {Gomboc}, {G{\'o}mez}, {Grux}, {Gueguen}, {Heyrovsky}, {Hoar}, {Iannicola},
  {Isasi Parache}, {Janotto}, {Joliet}, {Jonckheere}, {Keil}, {Kim},
  {Klagyivik}, {Klar}, {Knude}, {Kochukhov}, {Kolka}, {Kos}, {Kutka}, {Lainey},
  {LeBouquin}, {Liu}, {Loreggia}, {Makarov}, {Marseille}, {Martayan},
  {Martinez-Rubi}, {Massart}, {Meynadier}, {Mignot}, {Munari}, {Nguyen},
  {Nordlander}, {Ocvirk}, {O'Flaherty}, {Olias Sanz}, {Ortiz}, {Osorio},
  {Oszkiewicz}, {Ouzounis}, {Palmer}, {Park}, {Pasquato}, {Peltzer}, {Peralta},
  {P{\'e}turaud}, {Pieniluoma}, {Pigozzi}, {Poels}, {Prat}, {Prod'homme},
  {Raison}, {Rebordao}, {Risquez}, {Rocca-Volmerange}, {Rosen}, {Ruiz-Fuertes},
  {Russo}, {Sembay}, {Serraller Vizcaino}, {Short}, {Siebert}, {Silva},
  {Sinachopoulos}, {Slezak}, {Soffel}, {Sosnowska}, {Strai{\v{z}}ys}, {ter
  Linden}, {Terrell}, {Theil}, {Tiede}, {Troisi}, {Tsalmantza}, {Tur},
  {Vaccari}, {Vachier}, {Valles}, {Van Hamme}, {Veltz}, {Virtanen}, {Wallut},
  {Wichmann}, {Wilkinson}, {Ziaeepour}, \& {Zschocke}}]{Gaia_2016b}
{Gaia Collaboration}, {Prusti}, T., {de Bruijne}, J.~H.~J., {et~al.} 2016,
  \aap, 595, A1

\bibitem[{{Gaia Collaboration} {et~al.}(2022){Gaia Collaboration}, {Vallenari},
  {Brown}, {Prusti}, {de Bruijne}, {Arenou}, {Babusiaux}, {Biermann},
  {Creevey}, {Ducourant}, {Evans}, {Eyer}, {Guerra}, {Hutton}, {Jordi},
  {Klioner}, {Lammers}, {Lindegren}, {Luri}, {Mignard}, {Panem}, {Pourbaix},
  {Randich}, {Sartoretti}, {Soubiran}, {Tanga}, {Walton}, {Bailer-Jones},
  {Bastian}, {Drimmel}, {Jansen}, {Katz}, {Lattanzi}, {van Leeuwen}, {Bakker},
  {Cacciari}, {Casta{\~n}eda}, {De Angeli}, {Fabricius}, {Fouesneau},
  {Fr{\'e}mat}, {Galluccio}, {Guerrier}, {Heiter}, {Masana}, {Messineo},
  {Mowlavi}, {Nicolas}, {Nienartowicz}, {Pailler}, {Panuzzo}, {Riclet}, {Roux},
  {Seabroke}, {Sordo{\o}rcit}, {Th{\'e}venin}, {Gracia-Abril}, {Portell},
  {Teyssier}, {Altmann}, {Andrae}, {Audard}, {Bellas-Velidis}, {Benson},
  {Berthier}, {Blomme}, {Burgess}, {Busonero}, {Busso}, {C{\'a}novas}, {Carry},
  {Cellino}, {Cheek}, {Clementini}, {Damerdji}, {Davidson}, {de Teodoro},
  {Nu{\~n}ez Campos}, {Delchambre}, {Dell'Oro}, {Esquej},
  {Fern{\'a}ndez-Hern{\'a}ndez}, {Fraile}, {Garabato}, {Garc{\'\i}a-Lario},
  {Gosset}, {Haigron}, {Halbwachs}, {Hambly}, {Harrison}, {Hern{\'a}ndez},
  {Hestroffer}, {Hodgkin}, {Holl}, {Jan{\ss}en}, {Jevardat de Fombelle},
  {Jordan}, {Krone-Martins}, {Lanzafame}, {L{\"o}ffler}, {Marchal}, {Marrese},
  {Moitinho}, {Muinonen}, {Osborne}, {Pancino}, {Pauwels}, {Recio-Blanco},
  {Reyl{\'e}}, {Riello}, {Rimoldini}, {Roegiers}, {Rybizki}, {Sarro}, {Siopis},
  {Smith}, {Sozzetti}, {Utrilla}, {van Leeuwen}, {Abbas}, {{\'A}brah{\'a}m},
  {Abreu Aramburu}, {Aerts}, {Aguado}, {Ajaj}, {Aldea-Montero}, {Altavilla},
  {{\'A}lvarez}, {Alves}, {Anders}, {Anderson}, {Anglada Varela}, {Antoja},
  {Baines}, {Baker}, {Balaguer-N{\'u}{\~n}ez}, {Balbinot}, {Balog}, {Barache},
  {Barbato}, {Barros}, {Barstow}, {Bartolom{\'e}}, {Bassilana}, {Bauchet},
  {Becciani}, {Bellazzini}, {Berihuete}, {Bernet}, {Bertone}, {Bianchi},
  {Binnenfeld}, {Blanco-Cuaresma}, {Blazere}, {Boch}, {Bombrun}, {Bossini},
  {Bouquillon}, {Bragaglia}, {Bramante}, {Breedt}, {Bressan}, {Brouillet},
  {Brugaletta}, {Bucciarelli}, {Burlacu}, {Butkevich}, {Buzzi}, {Caffau},
  {Cancelliere}, {Cantat-Gaudin}, {Carballo}, {Carlucci}, {Carnerero},
  {Carrasco}, {Casamiquela}, {Castellani}, {Castro-Ginard}, {Chaoul},
  {Charlot}, {Chemin}, {Chiaramida}, {Chiavassa}, {Chornay}, {Comoretto},
  {Contursi}, {Cooper}, {Cornez}, {Cowell}, {Crifo}, {Cropper}, {Crosta},
  {Crowley}, {Dafonte}, {Dapergolas}, {David}, {David}, {de Laverny}, {De
  Luise}, {De March}, {De Ridder}, {de Souza}, {de Torres}, {del Peloso}, {del
  Pozo}, {Delbo}, {Delgado}, {Delisle}, {Demouchy}, {Dharmawardena}, {Di
  Matteo}, {Diakite}, {Diener}, {Distefano}, {Dolding}, {Edvardsson}, {Enke},
  {Fabre}, {Fabrizio}, {Faigler}, {Fedorets}, {Fernique}, {Fienga}, {Figueras},
  {Fournier}, {Fouron}, {Fragkoudi}, {Gai}, {Garcia-Gutierrez},
  {Garcia-Reinaldos}, {Garc{\'\i}a-Torres}, {Garofalo}, {Gavel}, {Gavras},
  {Gerlach}, {Geyer}, {Giacobbe}, {Gilmore}, {Girona}, {Giuffrida}, {Gomel},
  {Gomez}, {Gonz{\'a}lez-N{\'u}{\~n}ez}, {Gonz{\'a}lez-Santamar{\'\i}a},
  {Gonz{\'a}lez-Vidal}, {Granvik}, {Guillout}, {Guiraud},
  {Guti{\'e}rrez-S{\'a}nchez}, {Guy}, {Hatzidimitriou}, {Hauser}, {Haywood},
  {Helmer}, {Helmi}, {Sarmiento}, {Hidalgo}, {Hilger}, {H{\l}adczuk}, {Hobbs},
  {Holland}, {Huckle}, {Jardine}, {Jasniewicz}, {Jean-Antoine Piccolo},
  {Jim{\'e}nez-Arranz}, {Jorissen}, {Juaristi Campillo}, {Julbe}, {Karbevska},
  {Kervella}, {Khanna}, {Kontizas}, {Kordopatis}, {Korn}, {K{\'o}sp{\'a}l},
  {Kostrzewa-Rutkowska}, {Kruszy{\'n}ska}, {Kun}, {Laizeau}, {Lambert},
  {Lanza}, {Lasne}, {Le Campion}, {Lebreton}, {Lebzelter}, {Leccia}, {Leclerc},
  {Lecoeur-Taibi}, {Liao}, {Licata}, {Lindstr{\o}m}, {Lister}, {Livanou},
  {Lobel}, {Lorca}, {Loup}, {Madrero Pardo}, {Magdaleno Romeo}, {Managau},
  {Mann}, {Manteiga}, {Marchant}, {Marconi}, {Marcos}, {Marcos Santos},
  {Mar{\'\i}n Pina}, {Marinoni}, {Marocco}, {Marshall}, {Polo},
  {Mart{\'\i}n-Fleitas}, {Marton}, {Mary}, {Masip}, {Massari},
  {Mastrobuono-Battisti}, {Mazeh}, {McMillan}, {Messina}, {Michalik}, {Millar},
  {Mints}, {Molina}, {Molinaro}, {Moln{\'a}r}, {Monari}, {Mongui{\'o}},
  {Montegriffo}, {Montero}, {Mor}, {Mora}, {Morbidelli}, {Morel}, {Morris},
  {Muraveva}, {Murphy}, {Musella}, {Nagy}, {Noval}, {Oca{\~n}a}, {Ogden},
  {Ordenovic}, {Osinde}, {Pagani}, {Pagano}, {Palaversa}, {Palicio},
  {Pallas-Quintela}, {Panahi}, {Payne-Wardenaar}, {Pe{\~n}alosa Esteller},
  {Penttil{\"a}}, {Pichon}, {Piersimoni}, {Pineau}, {Plachy}, {Plum}, {Poggio},
  {Pr{\v{s}}a}, {Pulone}, {Racero}, {Ragaini}, {Rainer}, {Raiteri}, {Rambaux},
  {Ramos}, {Ramos-Lerate}, {Re Fiorentin}, {Regibo}, {Richards}, {Rios Diaz},
  {Ripepi}, {Riva}, {Rix}, {Rixon}, {Robichon}, {Robin}, {Robin}, {Roelens},
  {Rogues}, {Rohrbasser}, {Romero-G{\'o}mez}, {Rowell}, {Royer}, {Ruz Mieres},
  {Rybicki}, {Sadowski}, {S{\'a}ez N{\'u}{\~n}ez}, {Sagrist{\`a} Sell{\'e}s},
  {Sahlmann}, {Salguero}, {Samaras}, {Sanchez Gimenez}, {Sanna},
  {Santove{\~n}a}, {Sarasso}, {Schultheis}, {Sciacca}, {Segol}, {Segovia},
  {S{\'e}gransan}, {Semeux}, {Shahaf}, {Siddiqui}, {Siebert}, {Siltala},
  {Silvelo}, {Slezak}, {Slezak}, {Smart}, {Snaith}, {Solano}, {Solitro},
  {Souami}, {Souchay}, {Spagna}, {Spina}, {Spoto}, {Steele},
  {Steidelm{\"u}ller}, {Stephenson}, {S{\"u}veges}, {Surdej}, {Szabados},
  {Szegedi-Elek}, {Taris}, {Taylo}, {Teixeira}, {Tolomei}, {Tonello}, {Torra},
  {Torra}, {Torralba Elipe}, {Trabucchi}, {Tsounis}, {Turon}, {Ulla}, {Unger},
  {Vaillant}, {van Dillen}, {van Reeven}, {Vanel}, {Vecchiato}, {Viala},
  {Vicente}, {Voutsinas}, {Weiler}, {Wevers}, {Wyrzykowski}, {Yoldas}, {Yvard},
  {Zhao}, {Zorec}, {Zucker}, \& {Zwitter}}]{Gaia_2022K}
{Gaia Collaboration}, {Vallenari}, A., {Brown}, A.~G.~A., {et~al.} 2022, arXiv
  e-prints, arXiv:2208.00211

\bibitem[{{Gent} {et~al.}(2024){Gent}, {Eitner}, {Serenelli}, {Friske},
  {Koposov}, {Laporte}, {Buck}, \& {Bergemann}}]{Gent_2024}
{Gent}, M.~R., {Eitner}, P., {Serenelli}, A., {et~al.} 2024, \aap, 683, A74

\bibitem[{{Gilmore} {et~al.}(2012){Gilmore}, {Randich}, {Asplund}, {Binney},
  {Bonifacio}, {Drew}, {Feltzing}, {Ferguson}, {Jeffries}, {Micela},
  {Negueruela}, {Prusti}, {Rix}, {Vallenari}, {Alfaro}, {Allende-Prieto},
  {Babusiaux}, {Bensby}, {Blomme}, {Bragaglia}, {Flaccomio}, {Fran{\c{c}}ois},
  {Irwin}, {Koposov}, {Korn}, {Lanzafame}, {Pancino}, {Paunzen},
  {Recio-Blanco}, {Sacco}, {Smiljanic}, {Van Eck}, {Walton}, {Aden}, {Aerts},
  {Affer}, {Alcala}, {Altavilla}, {Alves}, {Antoja}, {Arenou}, {Argiroffi},
  {Asensio Ramos}, {Bailer-Jones}, {Balaguer-Nunez}, {Bayo}, {Barbuy},
  {Barisevicius}, {Barrado y Navascues}, {Battistini}, {Bellas Velidis},
  {Bellazzini}, {Belokurov}, {Bergemann}, {Bertelli}, {Biazzo}, {Bienayme},
  {Bland-Hawthorn}, {Boeche}, {Bonito}, {Boudreault}, {Bouvier}, {Brandao},
  {Brown}, {de Bruijne}, {Burleigh}, {Caballero}, {Caffau}, {Calura},
  {Capuzzo-Dolcetta}, {Caramazza}, {Carraro}, {Casagrande}, {Casewell},
  {Chapman}, {Chiappini}, {Chorniy}, {Christlieb}, {Cignoni}, {Cocozza},
  {Colless}, {Collet}, {Collins}, {Correnti}, {Covino}, {Crnojevic}, {Cropper},
  {Cunha}, {Damiani}, {David}, {Delgado}, {Duffau}, {Edvardsson}, {Eldridge},
  {Enke}, {Eriksson}, {Evans}, {Eyer}, {Famaey}, {Fellhauer}, {Ferreras},
  {Figueras}, {Fiorentino}, {Flynn}, {Folha}, {Franciosini}, {Frasca},
  {Freeman}, {Fremat}, {Friel}, {Gaensicke}, {Gameiro}, {Garzon}, {Geier},
  {Geisler}, {Gerhard}, {Gibson}, {Gomboc}, {Gomez}, {Gonzalez-Fernandez},
  {Gonzalez Hernandez}, {Gosset}, {Grebel}, {Greimel}, {Groenewegen},
  {Grundahl}, {Guarcello}, {Gustafsson}, {Hadrava}, {Hatzidimitriou}, {Hambly},
  {Hammersley}, {Hansen}, {Haywood}, {Heber}, {Heiter}, {Held}, {Helmi},
  {Hensler}, {Herrero}, {Hill}, {Hodgkin}, {Huelamo}, {Huxor}, {Ibata},
  {Jackson}, {de Jong}, {Jonker}, {Jordan}, {Jordi}, {Jorissen}, {Katz},
  {Kawata}, {Keller}, {Kharchenko}, {Klement}, {Klutsch}, {Knude}, {Koch},
  {Kochukhov}, {Kontizas}, {Koubsky}, {Lallement}, {de Laverny}, {van Leeuwen},
  {Lemasle}, {Lewis}, {Lind}, {Lindstrom}, {Lobel}, {Lopez Santiago}, {Lucas},
  {Ludwig}, {Lueftinger}, {Magrini}, {Maiz Apellaniz}, {Maldonado}, {Marconi},
  {Marino}, {Martayan}, {Martinez-Valpuesta}, {Matijevic}, {McMahon},
  {Messina}, {Meyer}, {Miglio}, {Mikolaitis}, {Minchev}, {Minniti}, {Moitinho},
  {Momany}, {Monaco}, {Montalto}, {Monteiro}, {Monier}, {Montes}, {Mora},
  {Moraux}, {Morel}, {Mowlavi}, {Mucciarelli}, {Munari}, {Napiwotzki},
  {Nardetto}, {Naylor}, {Naze}, {Nelemans}, {Okamoto}, {Ortolani}, {Pace},
  {Palla}, {Palous}, {Parker}, {Penarrubia}, {Pillitteri}, {Piotto}, {Posbic},
  {Prisinzano}, {Puzeras}, {Quirrenbach}, {Ragaini}, {Read}, {Read}, {Reyle},
  {De Ridder}, {Robichon}, {Robin}, {Roeser}, {Romano}, {Royer}, {Ruchti},
  {Ruzicka}, {Ryan}, {Ryde}, {Santos}, {Sanz Forcada}, {Sarro Baro},
  {Sbordone}, {Schilbach}, {Schmeja}, {Schnurr}, {Schoenrich}, {Scholz},
  {Seabroke}, {Sharma}, {De Silva}, {Smith}, {Solano}, {Sordo}, {Soubiran},
  {Sousa}, {Spagna}, {Steffen}, {Steinmetz}, {Stelzer}, {Stempels},
  {Tabernero}, {Tautvaisiene}, {Thevenin}, {Torra}, {Tosi}, {Tolstoy}, {Turon},
  {Walker}, {Wambsganss}, {Worley}, {Venn}, {Vink}, {Wyse}, {Zaggia},
  {Zeilinger}, {Zoccali}, {Zorec}, {Zucker}, {Zwitter}, \& {Gaia-ESO Survey
  Team}}]{Gilmore_2012}
{Gilmore}, G., {Randich}, S., {Asplund}, M., {et~al.} 2012, The Messenger, 147,
  25

\bibitem[{{Haywood}(2006)}]{Haywood_2006}
{Haywood}, M. 2006, \mnras, 371, 1760

\bibitem[{{Haywood}(2008)}]{Haywood_2008}
{Haywood}, M. 2008, \aap, 482, 673

\bibitem[{{Haywood} {et~al.}(2013){Haywood}, {Di Matteo}, {Lehnert}, {Katz}, \&
  {G{\'o}mez}}]{Haywood_2013}
{Haywood}, M., {Di Matteo}, P., {Lehnert}, M.~D., {Katz}, D., \& {G{\'o}mez},
  A. 2013, \aap, 560, A109

\bibitem[{{Haywood} {et~al.}(2016){Haywood}, {Lehnert}, {Di Matteo}, {Snaith},
  {Schultheis}, {Katz}, \& {G{\'o}mez}}]{Haywood_2016}
{Haywood}, M., {Lehnert}, M.~D., {Di Matteo}, P., {et~al.} 2016, \aap, 589, A66

\bibitem[{{Howes} {et~al.}(2019){Howes}, {Lindegren}, {Feltzing}, {Church}, \&
  {Bensby}}]{Howes_2019}
{Howes}, L.~M., {Lindegren}, L., {Feltzing}, S., {Church}, R.~P., \& {Bensby},
  T. 2019, \aap, 622, A27

\bibitem[{{Imig} {et~al.}(2023){Imig}, {Price}, {Holtzman}, {Stone-Martinez},
  {Majewski}, {Weinberg}, {Johnson}, {Allende Prieto}, {Beaton}, {Beers},
  {Bizyaev}, {Blanton}, {Brownstein}, {Cunha}, {Fern{\'a}ndez-Trincado},
  {Feuillet}, {Hasselquist}, {Hayes}, {J{\"o}nsson}, {Lane}, {Lian},
  {M{\'e}sz{\'a}ros}, {Nidever}, {Robin}, {Shetrone}, {Smith}, \&
  {Wilson}}]{Imig_2023}
{Imig}, J., {Price}, C., {Holtzman}, J.~A., {et~al.} 2023, \apj, 954, 124

\bibitem[{{Jofr{\'e}} {et~al.}(2020){Jofr{\'e}}, {Jackson}, \& {Tucci
  Maia}}]{Jofre_2020}
{Jofr{\'e}}, P., {Jackson}, H., \& {Tucci Maia}, M. 2020, \aap, 633, L9

\bibitem[{{Karakas} \& {Lattanzio}(2014)}]{Karakas_2014}
{Karakas}, A.~I. \& {Lattanzio}, J.~C. 2014, \pasa, 31, e030

\bibitem[{{Kim} {et~al.}(2002){Kim}, {Demarque}, {Yi}, \&
  {Alexander}}]{Kim_2002}
{Kim}, Y.-C., {Demarque}, P., {Yi}, S.~K., \& {Alexander}, D.~R. 2002, \apjs,
  143, 499

\bibitem[{{Kobayashi} {et~al.}(2006){Kobayashi}, {Umeda}, {Nomoto}, {Tominaga},
  \& {Ohkubo}}]{Kobayashi_2006}
{Kobayashi}, C., {Umeda}, H., {Nomoto}, K., {Tominaga}, N., \& {Ohkubo}, T.
  2006, \apj, 653, 1145

\bibitem[{{Lee} {et~al.}(2011){Lee}, {Beers}, {An}, {Ivezi{\'c}}, {Just},
  {Rockosi}, {Morrison}, {Johnson}, {Sch{\"o}nrich}, {Bird}, {Yanny},
  {Harding}, \& {Rocha-Pinto}}]{Lee_2011}
{Lee}, Y.~S., {Beers}, T.~C., {An}, D., {et~al.} 2011, \apj, 738, 187

\bibitem[{{Leung} {et~al.}(2023){Leung}, {Bovy}, {Mackereth}, \&
  {Miglio}}]{Leung_2023}
{Leung}, H.~W., {Bovy}, J., {Mackereth}, J.~T., \& {Miglio}, A. 2023, \mnras,
  522, 4577

\bibitem[{{Li} {et~al.}(2022){Li}, {Wang}, {Luo}, {Li}, {Deng}, \&
  {Ting}}]{Li_2022}
{Li}, Q.-D., {Wang}, H.-F., {Luo}, Y.-P., {et~al.} 2022, \apjs, 262, 20

\bibitem[{{Lin} {et~al.}(2020){Lin}, {Asplund}, {Ting}, {Casagrande}, {Buder},
  {Bland-Hawthorn}, {Casey}, {De Silva}, {D'Orazi}, {Freeman}, {Kos}, {Lind},
  {Martell}, {Sharma}, {Simpson}, {Zwitter}, {Zucker}, {Minchev},
  {{\v{C}}otar}, {Hayden}, {Horner}, {Lewis}, {Nordlander}, {Wyse}, \&
  {{\v{Z}}erjal}}]{Lin_2020}
{Lin}, J., {Asplund}, M., {Ting}, Y.-S., {et~al.} 2020, \mnras, 491, 2043

\bibitem[{{Liu} \& {van de Ven}(2012)}]{Liu_2012}
{Liu}, C. \& {van de Ven}, G. 2012, \mnras, 425, 2144

\bibitem[{{Martos} {et~al.}(2023){Martos}, {Mel{\'e}ndez}, {Rathsam}, \&
  {Carvalho Silva}}]{Martos_2023}
{Martos}, G., {Mel{\'e}ndez}, J., {Rathsam}, A., \& {Carvalho Silva}, G. 2023,
  \mnras, 522, 3217

\bibitem[{{Matteucci}(2014)}]{Matteucci_2014}
{Matteucci}, F. 2014, in Saas-Fee Advanced Course, Vol.~37, Saas-Fee Advanced
  Course, ed. J.~{Bland-Hawthorn}, K.~{Freeman}, \& F.~{Matteucci}, 145

\bibitem[{{Mel{\'e}ndez} {et~al.}(2012){Mel{\'e}ndez}, {Bergemann}, {Cohen},
  {Endl}, {Karakas}, {Ram{\'\i}rez}, {Cochran}, {Yong}, {MacQueen},
  {Kobayashi}, \& {Asplund}}]{Melendez_2012}
{Mel{\'e}ndez}, J., {Bergemann}, M., {Cohen}, J.~G., {et~al.} 2012, \aap, 543,
  A29

\bibitem[{{Mel{\'e}ndez} {et~al.}(2010){Mel{\'e}ndez}, {Casagrande},
  {Ram{\'\i}rez}, {Asplund}, \& {Schuster}}]{Melendez_2010}
{Mel{\'e}ndez}, J., {Casagrande}, L., {Ram{\'\i}rez}, I., {Asplund}, M., \&
  {Schuster}, W.~J. 2010, \aap, 515, L3

\bibitem[{{Mel{\'e}ndez} {et~al.}(2006){Mel{\'e}ndez}, {Dodds-Eden}, \&
  {Robles}}]{Melendez_2006}
{Mel{\'e}ndez}, J., {Dodds-Eden}, K., \& {Robles}, J.~A. 2006, \apjl, 641, L133

\bibitem[{{Mel{\'e}ndez} {et~al.}(2014){Mel{\'e}ndez}, {Ram{\'\i}rez},
  {Karakas}, {Yong}, {Monroe}, {Bedell}, {Bergemann}, {Asplund}, {Tucci Maia},
  {Bean}, {do Nascimento}, {Bazot}, {Alves-Brito}, {Freitas}, \&
  {Castro}}]{Melendez_2014}
{Mel{\'e}ndez}, J., {Ram{\'\i}rez}, I., {Karakas}, A.~I., {et~al.} 2014, \apj,
  791, 14

\bibitem[{{Moya} {et~al.}(2022){Moya}, {Sarro}, {Delgado-Mena}, {Chaplin},
  {Adibekyan}, \& {Blanco-Cuaresma}}]{Moya_2022}
{Moya}, A., {Sarro}, L.~M., {Delgado-Mena}, E., {et~al.} 2022, \aap, 660, A15

\bibitem[{{Navarro} {et~al.}(2011){Navarro}, {Abadi}, {Venn}, {Freeman}, \&
  {Anguiano}}]{Navarro_2011}
{Navarro}, J.~F., {Abadi}, M.~G., {Venn}, K.~A., {Freeman}, K.~C., \&
  {Anguiano}, B. 2011, \mnras, 412, 1203

\bibitem[{{Ness} {et~al.}(2019){Ness}, {Johnston}, {Blancato}, {Rix}, {Beane},
  {Bird}, \& {Hawkins}}]{Ness_2019}
{Ness}, M.~K., {Johnston}, K.~V., {Blancato}, K., {et~al.} 2019, \apj, 883, 177

\bibitem[{{Nidever} {et~al.}(2014){Nidever}, {Bovy}, {Bird}, {Andrews},
  {Hayden}, {Holtzman}, {Majewski}, {Smith}, {Robin}, {Garc{\'\i}a P{\'e}rez},
  {Cunha}, {Allende Prieto}, {Zasowski}, {Schiavon}, {Johnson}, {Weinberg},
  {Feuillet}, {Schneider}, {Shetrone}, {Sobeck}, {Garc{\'\i}a-Hern{\'a}ndez},
  {Zamora}, {Rix}, {Beers}, {Wilson}, {O'Connell}, {Minchev}, {Chiappini},
  {Anders}, {Bizyaev}, {Brewington}, {Ebelke}, {Frinchaboy}, {Ge}, {Kinemuchi},
  {Malanushenko}, {Malanushenko}, {Marchante}, {M{\'e}sz{\'a}ros}, {Oravetz},
  {Pan}, {Simmons}, \& {Skrutskie}}]{Nivedar_2014}
{Nidever}, D.~L., {Bovy}, J., {Bird}, J.~C., {et~al.} 2014, \apj, 796, 38

\bibitem[{{Nissen}(2015)}]{Nissen_2015}
{Nissen}, P.~E. 2015, \aap, 579, A52

\bibitem[{{Nissen}(2016)}]{Nissen_2016}
{Nissen}, P.~E. 2016, \aap, 593, A65

\bibitem[{{Nissen} {et~al.}(2020){Nissen}, {Christensen-Dalsgaard},
  {Mosumgaard}, {Silva Aguirre}, {Spitoni}, \& {Verma}}]{Nissen_2020}
{Nissen}, P.~E., {Christensen-Dalsgaard}, J., {Mosumgaard}, J.~R., {et~al.}
  2020, \aap, 640, A81

\bibitem[{{Nissen} {et~al.}(2017){Nissen}, {Silva Aguirre},
  {Christensen-Dalsgaard}, {Collet}, {Grundahl}, \& {Slumstrup}}]{Nissen_2017}
{Nissen}, P.~E., {Silva Aguirre}, V., {Christensen-Dalsgaard}, J., {et~al.}
  2017, \aap, 608, A112

\bibitem[{{Nomoto} {et~al.}(2013){Nomoto}, {Kobayashi}, \&
  {Tominaga}}]{Nomoto_2013}
{Nomoto}, K., {Kobayashi}, C., \& {Tominaga}, N. 2013, \araa, 51, 457

\bibitem[{{Patil} {et~al.}(2023){Patil}, {Bovy}, {Jaimungal}, {Frankel}, \&
  {Leung}}]{Patil_2023}
{Patil}, A.~A., {Bovy}, J., {Jaimungal}, S., {Frankel}, N., \& {Leung}, H.~W.
  2023, \mnras, 526, 1997

\bibitem[{{Ram{\'\i}rez} \& {Mel{\'e}ndez}(2005)}]{Ramirez_2005}
{Ram{\'\i}rez}, I. \& {Mel{\'e}ndez}, J. 2005, \apj, 626, 446

\bibitem[{{Ram{\'\i}rez} {et~al.}(2009){Ram{\'\i}rez}, {Mel{\'e}ndez}, \&
  {Asplund}}]{Ramirez_2009}
{Ram{\'\i}rez}, I., {Mel{\'e}ndez}, J., \& {Asplund}, M. 2009, \aap, 508, L17

\bibitem[{{Ram{\'\i}rez} {et~al.}(2014){Ram{\'\i}rez}, {Mel{\'e}ndez}, {Bean},
  {Asplund}, {Bedell}, {Monroe}, {Casagrande}, {Schirbel}, {Dreizler}, {Teske},
  {Tucci Maia}, {Alves-Brito}, \& {Baumann}}]{Ramirez_2014}
{Ram{\'\i}rez}, I., {Mel{\'e}ndez}, J., {Bean}, J., {et~al.} 2014, \aap, 572,
  A48

\bibitem[{{Randich} {et~al.}(2013){Randich}, {Gilmore}, \& {Gaia-ESO
  Consortium}}]{Randich_2013}
{Randich}, S., {Gilmore}, G., \& {Gaia-ESO Consortium}. 2013, The Messenger,
  154, 47

\bibitem[{{Rathsam} {et~al.}(2023){Rathsam}, {Mel{\'e}ndez}, \& {Carvalho
  Silva}}]{Rathsam_2023}
{Rathsam}, A., {Mel{\'e}ndez}, J., \& {Carvalho Silva}, G. 2023, \mnras, 525,
  4642

\bibitem[{{Reddy} {et~al.}(2006){Reddy}, {Lambert}, \& {Allende
  Prieto}}]{Reddy_2006}
{Reddy}, B.~E., {Lambert}, D.~L., \& {Allende Prieto}, C. 2006, \mnras, 367,
  1329

\bibitem[{{Rix} \& {Bovy}(2013)}]{Rix_2013}
{Rix}, H.-W. \& {Bovy}, J. 2013, \aapr, 21, 61

\bibitem[{{Sahlholdt} {et~al.}(2022){Sahlholdt}, {Feltzing}, \&
  {Feuillet}}]{Sahlholdt_2022}
{Sahlholdt}, C.~L., {Feltzing}, S., \& {Feuillet}, D.~K. 2022, \mnras, 510,
  4669

\bibitem[{{Salaris} {et~al.}(1993){Salaris}, {Chieffi}, \&
  {Straniero}}]{Salaris_1993}
{Salaris}, M., {Chieffi}, A., \& {Straniero}, O. 1993, \apj, 414, 580

\bibitem[{{Schirbel} {et~al.}(2015){Schirbel}, {Mel{\'e}ndez}, {Karakas},
  {Ram{\'\i}rez}, {Castro}, {Faria}, {Lugaro}, {Asplund}, {Tucci Maia}, {Yong},
  {Howes}, \& {do Nascimento}}]{Schirbel_2015}
{Schirbel}, L., {Mel{\'e}ndez}, J., {Karakas}, A.~I., {et~al.} 2015, \aap, 584,
  A116

\bibitem[{{Sch{\"o}nrich} \& {Binney}(2009)}]{Schonrich_2009}
{Sch{\"o}nrich}, R. \& {Binney}, J. 2009, \mnras, 396, 203

\bibitem[{{Sch{\"o}nrich} {et~al.}(2010){Sch{\"o}nrich}, {Binney}, \&
  {Dehnen}}]{Schonrich_2010}
{Sch{\"o}nrich}, R., {Binney}, J., \& {Dehnen}, W. 2010, \mnras, 403, 1829

\bibitem[{{Schuster} {et~al.}(2006){Schuster}, {Moitinho}, {M{\'a}rquez},
  {Parrao}, \& {Covarrubias}}]{Schuster_2006}
{Schuster}, W.~J., {Moitinho}, A., {M{\'a}rquez}, A., {Parrao}, L., \&
  {Covarrubias}, E. 2006, \aap, 445, 939

\bibitem[{{Shejeelammal} \& {Goswami}(2024)}]{Shejeelammal_2024}
{Shejeelammal}, J. \& {Goswami}, A. 2024, \mnras, 527, 2323

\bibitem[{{Silva Aguirre} {et~al.}(2012){Silva Aguirre}, {Casagrande}, {Basu},
  {Campante}, {Chaplin}, {Huber}, {Miglio}, {Serenelli}, {Ballot}, {Bedding},
  {Christensen-Dalsgaard}, {Creevey}, {Elsworth}, {Garc{\'\i}a}, {Gilliland},
  {Hekker}, {Kjeldsen}, {Mathur}, {Metcalfe}, {Monteiro}, {Mosser},
  {Pinsonneault}, {Stello}, {Weiss}, {Tenenbaum}, {Twicken}, \&
  {Uddin}}]{Silva_2012}
{Silva Aguirre}, V., {Casagrande}, L., {Basu}, S., {et~al.} 2012, \apj, 757, 99

\bibitem[{{Silva Aguirre} {et~al.}(2017){Silva Aguirre}, {Lund}, {Antia},
  {Ball}, {Basu}, {Christensen-Dalsgaard}, {Lebreton}, {Reese}, {Verma},
  {Casagrande}, {Justesen}, {Mosumgaard}, {Chaplin}, {Bedding}, {Davies},
  {Handberg}, {Houdek}, {Huber}, {Kjeldsen}, {Latham}, {White}, {Coelho},
  {Miglio}, \& {Rendle}}]{Silva_2017}
{Silva Aguirre}, V., {Lund}, M.~N., {Antia}, H.~M., {et~al.} 2017, \apj, 835,
  173

\bibitem[{{Sk{\'u}lad{\'o}ttir} {et~al.}(2019){Sk{\'u}lad{\'o}ttir}, {Hansen},
  {Salvadori}, \& {Choplin}}]{Skuladottir_2019}
{Sk{\'u}lad{\'o}ttir}, {\'A}., {Hansen}, C.~J., {Salvadori}, S., \& {Choplin},
  A. 2019, \aap, 631, A171

\bibitem[{{Slumstrup} {et~al.}(2017){Slumstrup}, {Grundahl}, {Brogaard},
  {Thygesen}, {Nissen}, {Jessen-Hansen}, {Van Eylen}, \&
  {Pedersen}}]{Slumstrup_2017}
{Slumstrup}, D., {Grundahl}, F., {Brogaard}, K., {et~al.} 2017, \aap, 604, L8

\bibitem[{{Sneden} {et~al.}(2012){Sneden}, {Bean}, {Ivans}, {Lucatello}, \&
  {Sobeck}}]{Sneden_2012}
{Sneden}, C., {Bean}, J., {Ivans}, I., {Lucatello}, S., \& {Sobeck}, J. 2012,
  {MOOG: LTE line analysis and spectrum synthesis}, Astrophysics Source Code
  Library, record ascl:1202.009

\bibitem[{{Sneden}(1973)}]{Sneden_1973}
{Sneden}, C.~A. 1973, PhD thesis, THE UNIVERSITY OF TEXAS AT AUSTIN.

\bibitem[{{Soderblom}(2010)}]{Soderblom_2010}
{Soderblom}, D.~R. 2010, \araa, 48, 581

\bibitem[{{Soderblom} \& {King}(1998)}]{Soderblom_1998}
{Soderblom}, D.~R. \& {King}, J.~R. 1998, in Solar Analogs: Characteristics and
  Optimum Candidates., ed. J.~C. {Hall}, 41

\bibitem[{{Spina} {et~al.}(2018){Spina}, {Mel{\'e}ndez}, {Karakas}, {dos
  Santos}, {Bedell}, {Asplund}, {Ram{\'\i}rez}, {Yong}, {Alves-Brito}, {Bean},
  \& {Dreizler}}]{Spina_2018}
{Spina}, L., {Mel{\'e}ndez}, J., {Karakas}, A.~I., {et~al.} 2018, \mnras, 474,
  2580

\bibitem[{{Spina} {et~al.}(2016){Spina}, {Mel{\'e}ndez}, {Karakas},
  {Ram{\'\i}rez}, {Monroe}, {Asplund}, \& {Yong}}]{Spina_2016b}
{Spina}, L., {Mel{\'e}ndez}, J., {Karakas}, A.~I., {et~al.} 2016, \aap, 593,
  A125

\bibitem[{{Tautvai{\v{s}}ien{\.{e}}} {et~al.}(2021){Tautvai{\v{s}}ien{\.{e}}},
  {Viscasillas V{\'a}zquez}, {Mikolaitis}, {Stonkut{\.{e}}},
  {Minkevi{\v{c}}i{\={u}}t{\.{e}}}, {Drazdauskas}, \&
  {Bagdonas}}]{Tautvaivsiene_2021}
{Tautvai{\v{s}}ien{\.{e}}}, G., {Viscasillas V{\'a}zquez}, C., {Mikolaitis},
  {\v{S}}., {et~al.} 2021, \aap, 649, A126

\bibitem[{{Titarenko} {et~al.}(2019){Titarenko}, {Recio-Blanco}, {de Laverny},
  {Hayden}, \& {Guiglion}}]{Titarenko_2019}
{Titarenko}, A., {Recio-Blanco}, A., {de Laverny}, P., {Hayden}, M., \&
  {Guiglion}, G. 2019, \aap, 622, A59

\bibitem[{{Tucci Maia} {et~al.}(2016){Tucci Maia}, {Ram{\'\i}rez},
  {Mel{\'e}ndez}, {Bedell}, {Bean}, \& {Asplund}}]{Tuccimaia_2016}
{Tucci Maia}, M., {Ram{\'\i}rez}, I., {Mel{\'e}ndez}, J., {et~al.} 2016, \aap,
  590, A32

\bibitem[{{Viscasillas V{\'a}zquez} {et~al.}(2022){Viscasillas V{\'a}zquez},
  {Magrini}, {Casali}, {Tautvai{\v{s}}ien{\.{e}}}, {Spina}, {Van der Swaelmen},
  {Randich}, {Bensby}, {Bragaglia}, {Friel}, {Feltzing}, {Sacco}, {Turchi},
  {Jim{\'e}nez-Esteban}, {D'Orazi}, {Delgado-Mena}, {Mikolaitis},
  {Drazdauskas}, {Minkevi{\v{c}}i{\={u}}t{\.{e}}}, {Stonkut{\.{e}}},
  {Bagdonas}, {Montes}, {Guiglion}, {Baratella}, {Tabernero}, {Gilmore},
  {Alfaro}, {Francois}, {Korn}, {Smiljanic}, {Bergemann}, {Franciosini},
  {Gonneau}, {Hourihane}, {Worley}, \& {Zaggia}}]{Viscasillas_2022}
{Viscasillas V{\'a}zquez}, C., {Magrini}, L., {Casali}, G., {et~al.} 2022,
  \aap, 660, A135

\bibitem[{{Wang} {et~al.}(2024){Wang}, {Carraro}, {Li}, {Li}, {Spina}, {Chen},
  {Wang}, \& {Deng}}]{Wang_2024}
{Wang}, H.-F., {Carraro}, G., {Li}, X., {et~al.} 2024, \apj, 967, 37

\bibitem[{{Weinberg} {et~al.}(2019){Weinberg}, {Holtzman}, {Hasselquist},
  {Bird}, {Johnson}, {Shetrone}, {Sobeck}, {Allende Prieto}, {Bizyaev},
  {Carrera}, {Cohen}, {Cunha}, {Ebelke}, {Fernandez-Trincado},
  {Garc{\'\i}a-Hern{\'a}ndez}, {Hayes}, {J{\"o}nsson}, {Lane}, {Majewski},
  {Malanushenko}, {M{\'e}sz{\'a}ros}, {Nidever}, {Nitschelm}, {Pan}, {Rix},
  {Rybizki}, {Schiavon}, {Schneider}, {Wilson}, \& {Zamora}}]{Weinberg_2019}
{Weinberg}, D.~H., {Holtzman}, J.~A., {Hasselquist}, S., {et~al.} 2019, \apj,
  874, 102

\bibitem[{{Xiang} \& {Rix}(2022)}]{Xiang_2022}
{Xiang}, M. \& {Rix}, H.-W. 2022, \nat, 603, 599

\bibitem[{{Yana Galarza} {et~al.}(2016){Yana Galarza}, {Mel{\'e}ndez},
  {Ram{\'\i}rez}, {Yong}, {Karakas}, {Asplund}, \& {Liu}}]{Galarza_2016}
{Yana Galarza}, J., {Mel{\'e}ndez}, J., {Ram{\'\i}rez}, I., {et~al.} 2016,
  \aap, 589, A17

\bibitem[{{Yanny} {et~al.}(2009){Yanny}, {Rockosi}, {Newberg}, {Knapp},
  {Adelman-McCarthy}, {Alcorn}, {Allam}, {Allende Prieto}, {An}, {Anderson},
  {Anderson}, {Bailer-Jones}, {Bastian}, {Beers}, {Bell}, {Belokurov},
  {Bizyaev}, {Blythe}, {Bochanski}, {Boroski}, {Brinchmann}, {Brinkmann},
  {Brewington}, {Carey}, {Cudworth}, {Evans}, {Evans}, {Gates}, {G{\"a}nsicke},
  {Gillespie}, {Gilmore}, {Nebot Gomez-Moran}, {Grebel}, {Greenwell}, {Gunn},
  {Jordan}, {Jordan}, {Harding}, {Harris}, {Hendry}, {Holder}, {Ivans},
  {Ivezi{\v{c}}}, {Jester}, {Johnson}, {Kent}, {Kleinman}, {Kniazev},
  {Krzesinski}, {Kron}, {Kuropatkin}, {Lebedeva}, {Lee}, {French Leger},
  {L{\'e}pine}, {Levine}, {Lin}, {Long}, {Loomis}, {Lupton}, {Malanushenko},
  {Malanushenko}, {Margon}, {Martinez-Delgado}, {McGehee}, {Monet}, {Morrison},
  {Munn}, {Neilsen}, {Nitta}, {Norris}, {Oravetz}, {Owen}, {Padmanabhan},
  {Pan}, {Peterson}, {Pier}, {Platson}, {Re Fiorentin}, {Richards}, {Rix},
  {Schlegel}, {Schneider}, {Schreiber}, {Schwope}, {Sibley}, {Simmons},
  {Snedden}, {Allyn Smith}, {Stark}, {Stauffer}, {Steinmetz}, {Stoughton},
  {SubbaRao}, {Szalay}, {Szkody}, {Thakar}, {Sivarani}, {Tucker}, {Uomoto},
  {Vanden Berk}, {Vidrih}, {Wadadekar}, {Watters}, {Wilhelm}, {Wyse}, {Yarger},
  \& {Zucker}}]{Yanny_2009}
{Yanny}, B., {Rockosi}, C., {Newberg}, H.~J., {et~al.} 2009, \aj, 137, 4377

\bibitem[{{Yi} {et~al.}(2001){Yi}, {Demarque}, {Kim}, {Lee}, {Ree}, {Lejeune},
  \& {Barnes}}]{Yi_2001}
{Yi}, S., {Demarque}, P., {Kim}, Y.-C., {et~al.} 2001, \apjs, 136, 417

\end{thebibliography}
\bibliographystyle{aa}

\clearpage
\onecolumn

\begin{appendix} 
\section{Stellar atmospheric parameters, mass, age, and elemental abundance ratios}

{\footnotesize
\begin{table*}[htp]
\centering
\caption{\small {Stellar atmospheric parameters, mass, age, and the elemental abundance ratios of the sample analysed in this study}.} \label{Table_results_1} 
\resizebox{0.7\textwidth}{!}{
\begin{tabular}{lccccccccccc} 
\hline                       
HD          & HIP                                       & T$_{eff}$  & log g  & [Fe/H]  & $\zeta$        & Age    & Mass   & [Mg/Fe]   & [Y/Fe]  & [Y/Mg]    \\  
            &                                           & (K)       & (dex)  & (dex)   & (km s$^{-1}$)    & (Gyr)  & (M$_{\odot}$) & (dex)  & (dex)  & (dex)  \\
\hline
208	      & 569     &  5929$\pm$7      &  4.45$\pm$0.02      &  $-$0.324$\pm$0.006      &  1.25$\pm$0.02      &  7.50$_{-0.24}^{+0.28}$      &  0.95$_{-0.01}^{+0.01}$      &  0.06$\pm$0.01      &  $-$0.08$\pm$0.01      &  $-$0.13$\pm$0.01	\\
4308	  & 3497    &  5688$\pm$6      &  4.40$\pm$0.02      &  $-$0.319$\pm$0.004      &  1.00$\pm$0.01      &  10.40$_{-0.24}^{+0.29}$      &  0.91$_{-0.02}^{+0.01}$      &  0.22$\pm$0.01      &  $-$0.12$\pm$0.01      &  $-$0.34$\pm$0.01	\\
4597	  & 3704    &  6027$\pm$12      &  4.36$\pm$0.03      &  $-$0.432$\pm$0.009      &  1.42$\pm$0.03      &  7.60$_{-0.25}^{+0.31}$      &  0.96$_{-0.01}^{+0.01}$      &  0.09$\pm$0.01      &  $-$0.13$\pm$0.01      &  $-$0.22$\pm$0.01	\\
7134	  & 5578    &  5992$\pm$9      &  4.52$\pm$0.02      &  $-$0.315$\pm$0.005      &  1.24$\pm$0.02      &  6.60$_{-0.25}^{+0.77}$      &  0.99$_{-0.02}^{+0.01}$      &  0.10$\pm$0.01      &  $-$0.03$\pm$0.01      &  $-$0.13$\pm$0.02	\\
17548	  & 13128   &  6004$\pm$14      &  4.36$\pm$0.04      &  $-$0.586$\pm$0.010      &  1.50$\pm$0.04      &  9.50$_{-0.32}^{+0.30}$      &  0.89$_{-0.01}^{+0.02}$      &  0.12$\pm$0.01      &  $-$0.16$\pm$0.01      &  $-$0.28$\pm$0.02	\\
17865	  & 13267   &  5882$\pm$11      &  4.25$\pm$0.03      &  $-$0.631$\pm$0.008      &  1.39$\pm$0.03      &  10.90$_{-0.28}^{+0.29}$      &  0.90$_{-0.01}^{+0.01}$      &  0.25$\pm$0.01      &  $-$0.11$\pm$0.01      &  $-$0.35$\pm$0.01	\\
\hline
\end{tabular} }
\end{table*} 
}

{\footnotesize
\begin{table*}[htp]
\centering
\caption{\small{Stellar atmospheric parameters, mass, age, and the elemental abundance ratios of the solar-twins and analogues}.} \label{Table_results_2} 
\resizebox{0.7\textwidth}{!}{\begin{tabular}{lccccccccccc} 
\hline                       
HD          & HIP                                & T$_{eff}$  & log g  & [Fe/H]  & $\zeta$        & Age    & Mass   & [Mg/Fe]   & [Y/Fe]  & [Y/Mg]    \\  
            &                                & (K)       & (dex)  & (dex)   & (km s$^{-1}$)    & (Gyr)  & (M$_{\odot}$) & (dex)  & (dex)  & (dex)  \\
\hline
361      & 669      &  5902$\pm$7      &  4.52$\pm$0.01      &  $-$0.122$\pm$0.005      &  1.10$\pm$0.01      &  1.60$_{-0.30}^{+0.42}$      &  1.02$_{-0.01}^{+0.02}$      &  $-$0.02$\pm$0.02      &  0.06$\pm$0.01      &  0.08$\pm$0.03	\\
1233     & 1292     &  5549$\pm$18      &  4.55$\pm$0.03      &  0.151$\pm$0.015      &  1.10$\pm$0.04      &  0.40$_{-0.20}^{+1.00}$      &  0.99$_{-0.02}^{+0.02}$      &  $-$0.07$\pm$0.01	      &  0.10$\pm$0.04      &  0.16$\pm$0.04	\\
1461	 & 1499     &  5751$\pm$4      &  4.33$\pm$0.01      &  0.182$\pm$0.003      &  1.01$\pm$0.01      &  4.70$_{-0.35}^{+0.36}$      &  1.07$_{-0.01}^{+0.01}$      &  0.00$\pm$0.01      &  $-$0.03$\pm$0.01      &  $-$0.03$\pm$0.02	\\
2071     & 1954	    &  5720$\pm$2      &  4.46$\pm$0.01      &  $-$0.090$\pm$0.003      &  0.96$\pm$0.01      &  4.80$_{-0.25}^{+0.39}$      &  0.97$_{-0.01}^{+0.01}$      &  0.01$\pm$0.01      &  0.03$\pm$0.01      &  0.02$\pm$0.01	\\
3821     & 3203	     &  5868$\pm$9      &  4.54$\pm$0.02      &  $-$0.050$\pm$0.007      &  1.16$\pm$0.02      &  0.53$_{-0.34}^{+0.34}$      &  1.04$_{-0.02}^{+0.01}$      &  $-$0.04$\pm$0.01      &  0.10$\pm$0.01      &  0.14$\pm$0.02	\\
3964     & 3311	     &  5734$\pm$3      &  4.47$\pm$0.01      &  0.064$\pm$0.003      &  0.96$\pm$0.01      &  3.40$_{-0.44}^{+0.25}$      &  1.01$_{-0.01}^{+0.01}$      &  $-$0.01$\pm$0.01      &  0.04$\pm$0.01      &  0.05$\pm$0.01	\\
\hline
\end{tabular} }
\end{table*} 
}

{\footnotesize
\begin{table*}[ht!]
\centering
\ContinuedFloat
\caption{continued} 
\resizebox{0.7\textwidth}{!}{\begin{tabular}{lccccccccccc} 
\hline                       
HD             & HIP                           & T$_{eff}$  & log g  & [Fe/H]  & $\zeta$        & Age    & Mass   & [Mg/Fe]   & [Y/Fe]  & [Y/Mg]    \\  
               &                            & (K)       & (dex)  & (dex)   & (km s$^{-1}$)    & (Gyr)  & (M$_{\odot}$) & (dex)  & (dex)  & (dex)  \\
\hline
16297    & 12119      &  5396$\pm$6      &  4.43$\pm$0.02      &  0.008$\pm$0.006      &  0.81$\pm$0.02      &  6.80$_{-0.46}^{+0.92}$      &  0.89$_{-0.01}^{+0.02}$      &  $-$0.02$\pm$0.01      &  0.07$\pm$0.02      &  0.09$\pm$0.03	\\
17925    & 13402      &  5149$\pm$18      &  4.44$\pm$0.06      &  0.088$\pm$0.015      &  0.97$\pm$0.06      &  6.10$_{-1.89}^{+2.07}$      &  0.84$_{-0.01}^{+0.01}$      &  $-$0.09$\pm$0.02      &  0.13$\pm$0.06      &  0.22$\pm$0.06	\\
19467    & 14501      &  5738$\pm$4      &  4.31$\pm$0.01      &  $-$0.153$\pm$0.003      &  1.07$\pm$0.01      &  9.80$_{-0.25}^{+0.19}$      &  0.98$_{-0.01}^{+0.01}$      &  0.17$\pm$0.01      &  $-$0.08$\pm$0.01      &  $-$0.25$\pm$0.01	\\
19518    & 14614      &  5803$\pm$4      &  4.45$\pm$0.01      &  $-$0.109$\pm$0.004      &  1.03$\pm$0.01      &  5.60$_{-0.47}^{+0.38}$      &  0.99$_{-0.01}^{+0.01}$      &  0.01$\pm$0.01      &  0.02$\pm$0.01      &  0.01$\pm$0.01	\\
19668    & 14684      &  5501$\pm$18      &  4.57$\pm$0.03      &  0.010$\pm$0.013      &  1.15$\pm$0.04      &  1.00$_{-0.40}^{+1.15}$      &  0.93$_{-0.01}^{+0.01}$      &  $-$0.08$\pm$0.02      &  0.09$\pm$0.05      &  0.18$\pm$0.06	\\
20003    & 14530      &  5483$\pm$5      &  4.40$\pm$0.02      &  0.075$\pm$0.006      &  0.76$\pm$0.02      &  7.60$_{-0.72}^{+0.26}$      &  0.93$_{-0.01}^{+0.01}$      &  0.00$\pm$0.01      &  0.04$\pm$0.02      &  0.03$\pm$0.02	\\
\hline
\end{tabular} }
\end{table*} 
}

{\footnotesize
\begin{table*}
\centering
\ContinuedFloat
\caption{continued}
\resizebox{0.7\textwidth}{!}{\begin{tabular}{lccccccccccc} 
\hline                       
HD            & HIP                            & T$_{eff}$  & log g  & [Fe/H]  & $\zeta$        & Age    & Mass   & [Mg/Fe]   & [Y/Fe]  & [Y/Mg]    \\  
              &                              & (K)       & (dex)  & (dex)   & (km s$^{-1}$)    & (Gyr)  & (M$_{\odot}$) & (dex)  & (dex)  & (dex)  \\
\hline
115169  & 64713       &  5788$\pm$4      &  4.44$\pm$0.01      &  $-$0.043$\pm$0.003      &  1.00$\pm$0.01      &  5.40$_{-0.41}^{+0.40}$      &  0.99$_{-0.01}^{+0.02}$      &  0.01$\pm$0.00      &  0.01$\pm$0.01      &  0.00$\pm$0.01	\\
115617   & 64924      &  5568$\pm$4      &  4.39$\pm$0.01      &  0.006$\pm$0.004      &  0.84$\pm$0.01      &  7.70$_{-0.26}^{+0.28}$      &  0.93$_{-0.01}^{+0.01}$      &  0.03$\pm$0.01      &  $-$0.06$\pm$0.02      &  $-$0.08$\pm$0.02	\\
117126   & 65708      &  5746$\pm$5      &  4.22$\pm$0.01      &  $-$0.063$\pm$0.005      &  1.12$\pm$0.01      &  9.10$_{-0.24}^{+0.25}$      &  1.01$_{-0.01}^{+0.02}$      &  0.11$\pm$0.01      &  $-$0.08$\pm$0.01      &  $-$0.19$\pm$0.01	\\
117207   & 65808      &  5659$\pm$7      &  4.30$\pm$0.02      &  0.242$\pm$0.007      &  0.97$\pm$0.02      &  5.40$_{-0.26}^{+0.29}$      &  1.05$_{-0.01}^{+0.02}$      &  $-$0.01$\pm$0.02      &  $-$0.04$\pm$0.03      &  $-$0.03$\pm$0.03	\\
122194   & 68468      &  5845$\pm$5      &  4.33$\pm$0.01      &  0.071$\pm$0.004      &  1.10$\pm$0.01      &  5.80$_{-0.25}^{+0.29}$      &  1.06$_{-0.01}^{+0.01}$      &  0.05$\pm$0.01      &  $-$0.05$\pm$0.01      &  $-$0.10$\pm$0.01	\\
124292   & 69414       &  5458$\pm$4      &  4.39$\pm$0.01      &  $-$0.113$\pm$0.004      &  0.79$\pm$0.01      &  9.90$_{-0.46}^{+0.24}$      &  0.87$_{-0.01}^{+0.01}$      &  0.06$\pm$0.01      &  $-$0.08$\pm$0.02      &  $-$0.13$\pm$0.02	\\
\hline
\end{tabular} }
\end{table*} 
}

\twocolumn

\section{Kinematics of the stellar sample}
\begin{figure}
\centering
\includegraphics[width=\columnwidth]{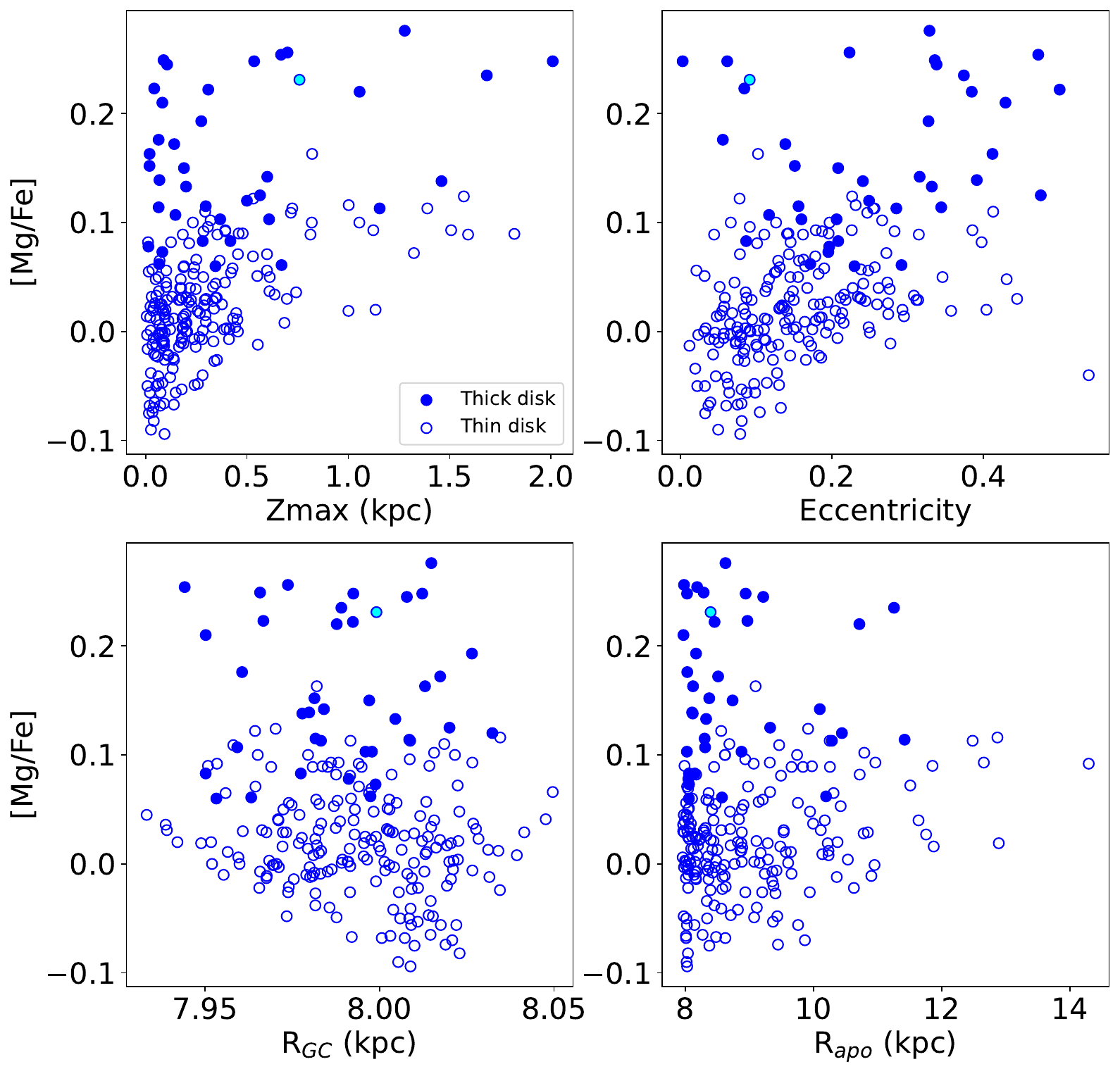} 
\caption{Observed [Mg/Fe] of the entire sample with respect to various orbital parameters. 
The cyan circle is the anomalous star HD 65907.}\label{MgFe_orbital_parameters}
\end{figure}

\begin{figure}
\centering
\includegraphics[width=\columnwidth]{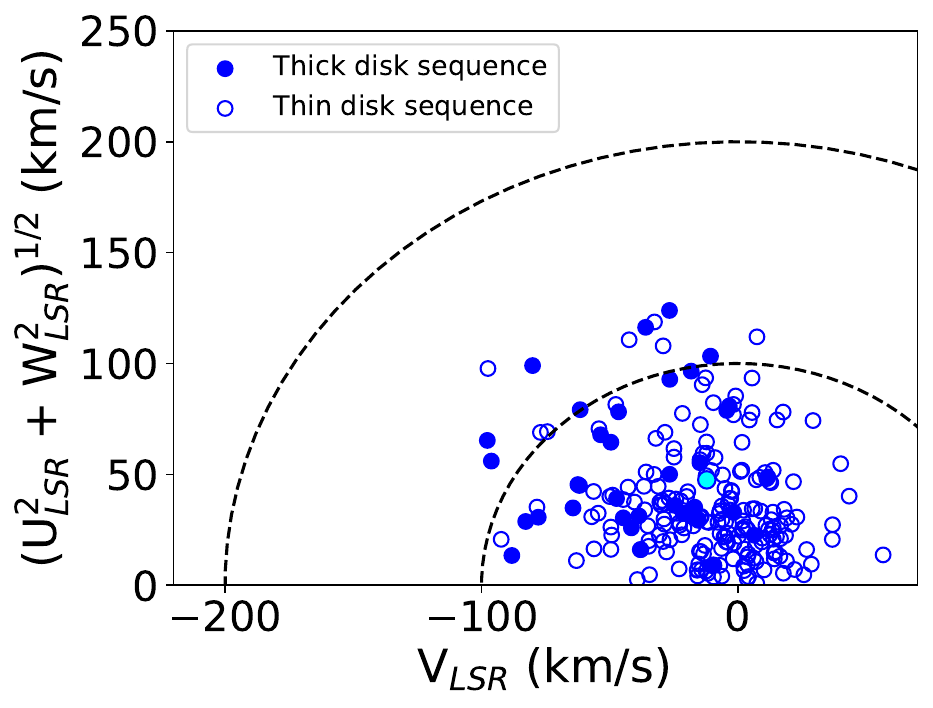} 
\caption{Toomre diagram for the entire sample. The dotted inner and outer circles 
have radii of 100 km/s and 200 km/s, respectively. }\label{Toomre_diagram}
\end{figure}

Various orbital characteristics and  Galactic space velocities of the sample are calculated using the 
\texttt{galpy} \footnote{\url{http://github.com/jobovy/galpy}} Python package \citep{Bovy_2015}.
The components of spatial velocities (U, V, and W) are calculated with respect to the local standard of rest (LSR). 
The adopted solar motion about the LSR is (U, V, W)$_{\odot}$ = (11.1, 12.2, 7.3) km/s \citep{Schonrich_2010}. 
The detailed procedure can be found in \cite{Shejeelammal_2024}. The behaviour of the [Mg/Fe] ratio observed in the 
sample as functions of maximum height above the Galactic plane (Z$_{max}$), eccentricity, Galactocentric distance (R$_{GC}$) and apocentric distance (R$_{apo}$)
are shown in Fig. \ref{MgFe_orbital_parameters}. The Toomre diagram for the sample is shown in Fig. \ref{Toomre_diagram}. 
As can be seen, it is difficult to separate the thin and thick disk from purely kinematic criteria.

\end{appendix}

\end{document}